\documentclass[superscriptaddress,longbibliography,aps,prl,reprint,showpacs]{revtex4-1}

\usepackage[english]{babel}
\usepackage[T1]{fontenc}
\usepackage{graphicx}
\usepackage{amsmath}
\usepackage{soul}
\usepackage{natbib}
\usepackage{eqnarray}
\usepackage{textcomp}

\usepackage{xcolor}

\bibstyle{aapmrev4-1}

\begin{document}
\title{Ultrafast negative thermal expansion driven by spin-disorder}
%
%
\author{J. Pudell}
\affiliation{Institut f\"ur Physik \& Astronomie,  Universit\"at Potsdam, Karl-Liebknecht-Str. 24-25, 14476 Potsdam,  Germany}

\author{A. von Reppert}
\affiliation{Institut f\"ur Physik \& Astronomie,  Universit\"at Potsdam, Karl-Liebknecht-Str. 24-25, 14476 Potsdam,  Germany}

\author{D. Schick}
\affiliation{Helmholtz Zentrum Berlin, Albert-Einstein-Str. 15, 12489 Berlin, Germany}

\author{F. Zamponi}
\affiliation{Institut f\"ur Physik \& Astronomie,  Universit\"at Potsdam, Karl-Liebknecht-Str. 24-25, 14476 Potsdam,  Germany}

\author{M. R\"ossle}	
\affiliation{Institut f\"ur Physik \& Astronomie,  Universit\"at Potsdam, Karl-Liebknecht-Str. 24-25, 14476 Potsdam,  Germany}
\affiliation{Helmholtz Zentrum Berlin, Albert-Einstein-Str. 15, 12489 Berlin, Germany}
%
%
%
\author{M. Herzog}
\affiliation{Institut f\"ur Physik \& Astronomie,  Universit\"at Potsdam, Karl-Liebknecht-Str.~24-25, 14476 Potsdam,  Germany}

\author{H. Zabel}
\affiliation{Fakult\"at f\"ur Physik und Astronomie, Ruhr-Universit\"at Bochum, 44780 Bochum, Germany}
\author{M. Bargheer}
\email{bargheer@uni-potsdam.de}
\homepage{http://www.udkm.physik.uni-potsdam.de} \affiliation{Institut  f\"ur Physik \& Astronomie, Universit\"at Potsdam,  Karl-Liebknecht-Str. 24-25, 14476 Potsdam, Germany}
\affiliation{Helmholtz Zentrum Berlin, Albert-Einstein-Str.~15, 12489 Berlin, Germany}

\newcommand{\superscript}[1]{\ensuremath{^{\textrm{#1}}}}
\newcommand{\subscript}[1]{\ensuremath{_{\textrm{#1}}}}

\date{\today}
\begin{abstract}
    We measure the transient strain profile in a nanoscale multilayer system composed of Yttrium, Holmium and Niobium after laser excitation using ultrafast X-ray diffraction. The strain propagation through each layer is determined by transient changes of the material-specific Bragg angles. We experimentally derive the exponentially decreasing stress profile driving the strain wave and show that it closely matches the optical penetration depth. Below the Neel temperature of Ho, the optical excitation 
    triggers negative thermal expansion, which is induced by  a quasi-instantaneous contractive stress, and a second contractive stress contribution rising on a 12\,ps timescale. These two timescales have recently been measured for the spin-disordering in Ho [Rettig et al, PRL 116, 257202 (2016)]. As a consequence we observe an unconventional bipolar strain pulse with an inverted sign travelling through the heterostructure.
\end{abstract}
\maketitle
In most of the research on ultrafast magnetism the lattice was only considered as an angular momentum sink.\cite{beau1996a,bigo2009,stam2007} Ultrafast effects on the lattice induced by demagnetization have been discussed surprisingly rarely. \cite{korf2008,reid2018,mali2008,rudo2012} Time-resolved magneto-optical Kerr measurements and optical picosecond ultrasonics are the workhorse for many researchers.\cite{thom1986a,beau1996a,koop2010a,hofh2017,bigo2009,kime2005,kim2012} Specialized techniques allow for assigning timescales to specific electronic processes and orbitals or bands.\cite{rett2016,stam2007,thie2017,pfau2012,doer2005} This is particularly relevant in the magnetic rare earths, where the exchange interaction between the localized $4f$ spin and orbital magnetic moments is mediated by the itinerant $5d6s$ conduction electrons via the RKKY interaction.\cite{rude1954,darn1963} 
Ultrafast electron diffraction (UED) or ultrafast x-ray diffraction (UXRD) experiments that directly observe the transient lattice strain induced by ultrafast demagnetization have been discussed only sporadically. \cite{korf2008,reid2018,koc2017,repp2016,quir2012}
Several ultrafast diffraction studies on the transition metals Ni and Fe \cite{wang2008, wang2010,heni2016} discuss the strain waves excited by electron and phonon stresses $\sigma_\text{e}$ and $\sigma_\text{ph}$, and theory predicts relevant electron-phonon (e-ph) coupling constants \cite{lin2008a} even with mode-specificity. \cite{mald2017}  Very recently granular FePt films were studied by UED. The rapid out-of-plane lattice contraction could be convincingly ascribed to changes of the free energy of the spin system. \cite{reid2018} The macroscopic Gr\"uneisen coefficients (Gc) $\Gamma_\text{e,ph}$ describe the efficiency for generating stress $\sigma_\text{e,ph}=\Gamma_\text{e,ph} \rho^Q_\text{e,ph}$ by a heat energy density $\rho^Q_\text{e,ph}$. \cite{barr2005} If $ \Gamma_\text{e} \neq \Gamma_\text{ph}$, ultrafast diffraction allows inferring the time-dependent $\sigma(t)$ from the observed transient strain $\varepsilon(t)$.\cite{wang2008, wang2010,nico2011a}  Hooke's law relates $\varepsilon$ linearly to  $\sigma$ and hence  to the energy densities $\rho^Q_\text{e,ph}$ deposited in each subsystem. This concept was extended to stress resulting from spin-excitations in Ni and Fe \cite{whit1962,barr1980} but the experimental verification remained ambiguous. \cite{wang2008, wang2010,heni2016}  Thermodynamic analysis affirms that the Gcs generally measure how entropy $S$ depends on strain $\varepsilon$. \cite{whit1989} The phenomenon of negative thermal expansion (NTE) generally occurs when the $S$ decreases upon expansion, i.e. $\partial S / \partial \varepsilon < 0$.   For spin-ordered phases of rare earths, the NTE along the $c$-axis of the hexagonal lattice is very large (cf. Fig. 1a-d). Surprisingly, also in these systems exhibiting a divergent specific heat around the second order phase transition to the paramagnetic (PM) phase, the spin-Gc \cite{repp2016,koc2017} $\Gamma_\text{sp}$ is essentially independent of $T$, even though the total Gc \cite{whit1989} varies strongly with $T$. Separating phonon and spin contributions to the thermal expansion and the heat capacity of Ho yields $\Gamma_{\text{ph}}/\Gamma_{\text{sp}}\approx -0.2$. \cite{whit1989,barr1980} The separate Gc $\Gamma_{\text{e,ph,sp}}$ are independent of $T$ because the $T$ dependence of the specific heat $C_{\text{e,ph,sp}}(T)$ and the thermal expansion coefficient $\alpha_{e,ph,sp}(T)$ cancel out.\cite{barr1980,barr2005}  Rare earth elements prove to be a versatile testing ground for understanding how rapidly ultrafast demagnetization leads to stress in the crystal lattice. We selected Ho since a recent resonant hard x-ray scattering experiment measured that the demagnetization of both the localized $4f$ moments and the itinerant conduction electrons proceed on a fast 200\,fs timescale attributed to electron-spin interaction and a slow 9\,ps timescale for coupling phonons to the spins. \cite{rett2016}

In this letter we use UXRD at a laser-based femtosecond x-ray plasma source (PXS) to show that the ultrafast laser excitation of Ho below its Neel temperature $T_N=132$\,K generates negative stress that rises on the two timescales for disordering the spin-system and drives bipolar strain waves with opposite polarity compared to common materials without NTE.\cite{thom1986a,schi2014c} We use the material-specificity of UXRD to cross-check the individual lattice constant changes in this metallic multilayer system, which is opaque to optical probes. In the PM phase of Ho, the analysis is simplified by the complete spin disorder and we show that the spatial excitation profile of the driving stress can be derived by probing the bipolar strain pulse in the adjacent Y and Nb buffer layers. 
The spatial stress profile is approximately given by the penetration depth of the pump-pulses. The UXRD experiment in the antiferromagnetic (AFM) phase at 35\,K reveals an instantaneous compensation of the expansive electron and phonon stress in Ho by the negative stress due to spin disorder. In addition to this sub-ps negative stress component, the negative stress keeps rising on a 12\,ps timescale. We observe  the extended leading edge of the bipolar strain pulse with opposite sign compared to the PM phase in the adjacent Y and Nb. The observed lattice contraction in Ho is about twenty times larger than the peak shift of the magnetic Bragg peak observed for very similar fluence, which in fact measures the incommensurate spin helix.\cite{rett2016} In a broader context, our analysis reveals, how fast the entropy driven NTE phenomenon can occur in various materials ranging from nonmagnetic molecular nanocrystals \cite{van2013} and oxides with open oxygen frameworks \cite{erns1998}, to ferroelectrics \cite{chen2013} and magnetically ordered systems like Heusler alloys.\cite{khme2003}

\begin{figure}
 \centering
		\includegraphics[width = 8cm]{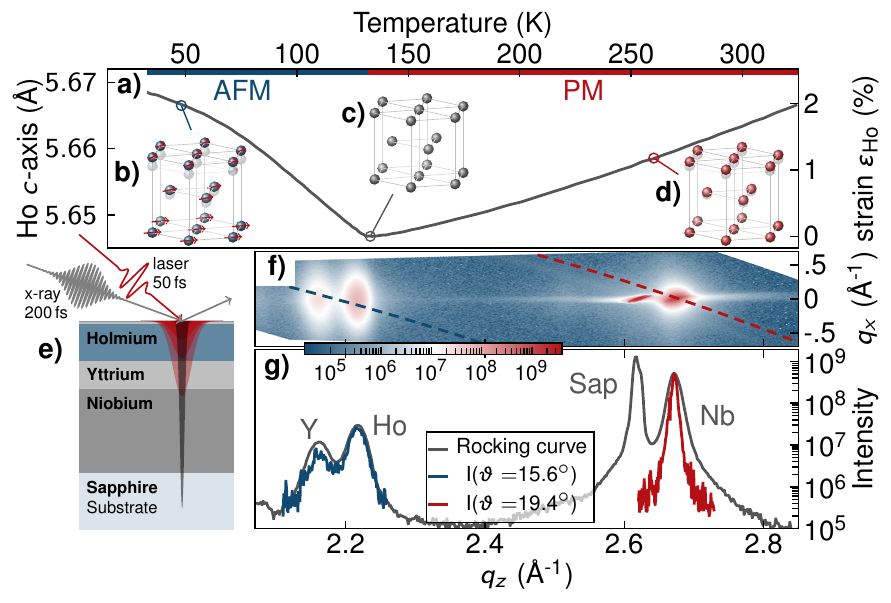}
	\caption{a) $c$-axis of the Ho thin film as a function of temperature, showing large NTE below $T_N=132$K. b-d) schematics of the hexagonal lattice illustrating the $c$-axis lattice change and the helical spin order below $T_N$. e) Schematic of  the layer stacking and the pump-probe geometry. f) The RSM including seperated Bragg peaks of Y, Ho, Sapphire and Nb. The slices of the RSM used for panel g) and the time-resolved measurements are shown in red (Nb) and blue (Ho+Y). g) shows the projection of the full RSM (grey) and the single angle measurements along the slices (red and blue).}
	\label{fig:static}
\end{figure}
The multilayer stack was grown by molecular beam epitaxy on sapphire in the sequence 128\,nm Nb, 34\,nm Y, 46\,nm Ho with a thin capping layer of 4\,nm Y and 3\,nm Nb (cf. Fig. 1e). 
200\,fs hard X-ray probe pulses with a photon energy of 8\,keV are derived from the PXS at the University of Potsdam. \cite{schi2012} P-polarized 50\,fs laser pulses with a diameter of 1500\,\textmu m (FWHM) excite the sample at an incidence angle of about 52$^\circ$. The incident fluence of the 800\,nm pump pulses is 3\,mJ/cm$^2$ an absorbed fluence of 1.7\,mJ/cm$^2$ is calculated according to the refractive index obtained by spectroscopic ellipsometry. The penetration depth is 21\,nm at 800\,nm for all temperatures.

The reciprocal space map (RSM) of the multilayer system including separated Bragg peaks of Y, Ho, sapphire and Nb at room temperature is shown in Fig.~\ref{fig:static}f). 
The grey line in Fig.~\ref{fig:static}f) shows a projection of the full RSM onto the out-of-plane component $q_z$. The red and blue lines indicate typical x-ray diffraction curves derived from the dashed cuts through reciprocal space indicated by the red and blue lines in panel f). 
\begin{figure}
   \centering
 \includegraphics[width = 8 cm]{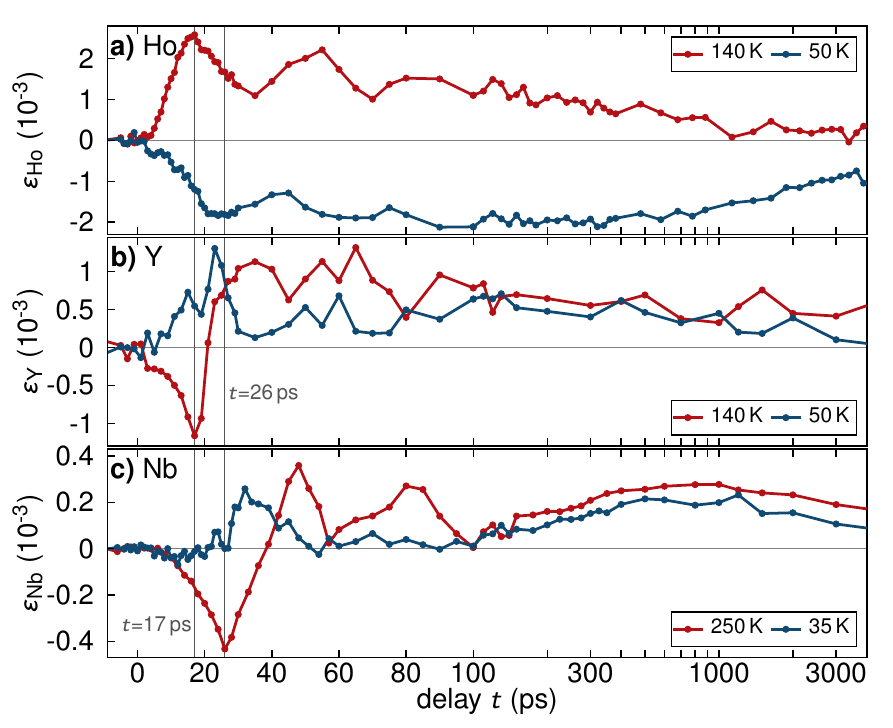}
  \caption{Transient lattice strain $\varepsilon$ for  a) Ho, b) Y and c) Nb after laser excitation in the PM (red) and AFM (blue) phase. In the AFM phase Ho shows a 6\,ps delay of the contraction compared to the PM expansion.}
\label{fig:data}
\end{figure}
Fig.~\ref{fig:data} summarizes the strain in all three layers for the PM phase of Ho (red) and in the helical AFM phase of Ho around 40\,K (blue).  
In the PM phase two characteristic times are identified from the UXRD data: At $t=17$\,ps, Ho has reached the maximum expansion. This indicates the time it takes the expansive sound to travel from the surface to the Y interface. At this time the leading compressive part of the bipolar strain wave\cite{thom1986a, schi2014c} has fully propagated from the Ho into the Y layer, as evidenced by the pronounced minimum in the Y strain (see Fig.~3 for an illustration of the strain wave). At $t=26$\,ps the zero crossing of the bipolar strain wave traverses the Y/Nb interface, which yields the pronounced minimum in the Nb strain. These characteristic time points give the layer thicknesses given above.

The UXRD data recorded in the AFM phase of Ho directly show a delay of the contractive strain in Ho (Fig.~2a, blue line). The minimum of the Ho strain at about 23\,ps is delayed by about 6\,ps compared to the maximum in the PM phase. The signals in Y (Fig.~2b) confirms, that also the propagating bipolar strain wave has the reversed sign of the strain amplitude at low temperature and a delay of about 6\,ps. In the PM phase, the leading compressive part of the strain launched in Ho by thermal expansion reduces the out-of plane lattice spacing in Y, although it is also excited by the pump pulse.\cite{schi2014c} In the AFM phase the contractive stress in Ho dominates and reverses the sign of the bipolar strain. Hence the Ho contraction expands the adjacent Y, assisted by the small expansive stress from direct optical excitation of Y. The UXRD signal from the Nb layer is even cleaner, as a negligible amount of light is absorbed in this layer. Therefore, we repeated the experiments on Nb. The obtained signal (Fig.~2c) had to be scaled appropriately due to the slightly different base temperature and fluence. The tiny negative lattice constant change of Nb (Fig.~2c) in the first 15\,ps is due to the stress generated in Y, and the crossover to the strong expansion starts around 10 ps, when the leading expansive part of the bipolar strain wave generated by the exponentially decaying compressive stress in Ho reaches the Nb layer. The very short wiggle of the average Nb layer strain at 26\,ps heralds the rather strong bipolar strain wave launched by the 7\,nm thick Nb/Y capping layer. The time perfectly coincides with the arrival of the wave created by surface expansion that was already observed in the PM phase. The same feature from the capping layer is also clearly observed in the Y data at 19\,ps. In the PM data set, this feature is absent since the cap layer and the Ho layer both expand with similar amplitude.

For $t>200$\,ps the sound wave reflections have ceased and we can safely interpret the expansion and contraction of each layer by the average energy densities $\rho_\text{e,ph,s}$. 
Before that the measured transient (negative) thermal expansion $\varepsilon(t) = \varepsilon_\text{th}(t)+\varepsilon_\text{sw}(t)$ is a linear superposition of the averaged thermal strain $\varepsilon_\text{th}(t)$ and the layer-averaged amplitude of the hypersound waves $\varepsilon_\text{sw}(t)$  triggered by the rapid expansion following the ultrashort pulse excitation, which reflect from the interfaces.  
Thus, the temporal stress can be obtained from the data by averaging out the oscillations. For timescales shorter than the characteristic oscillation period of the layer, however, the transient stress must be obtained from modelling the elastic response of the system.
\begin{figure}
   \centering
 \includegraphics[width = 8.
  cm]{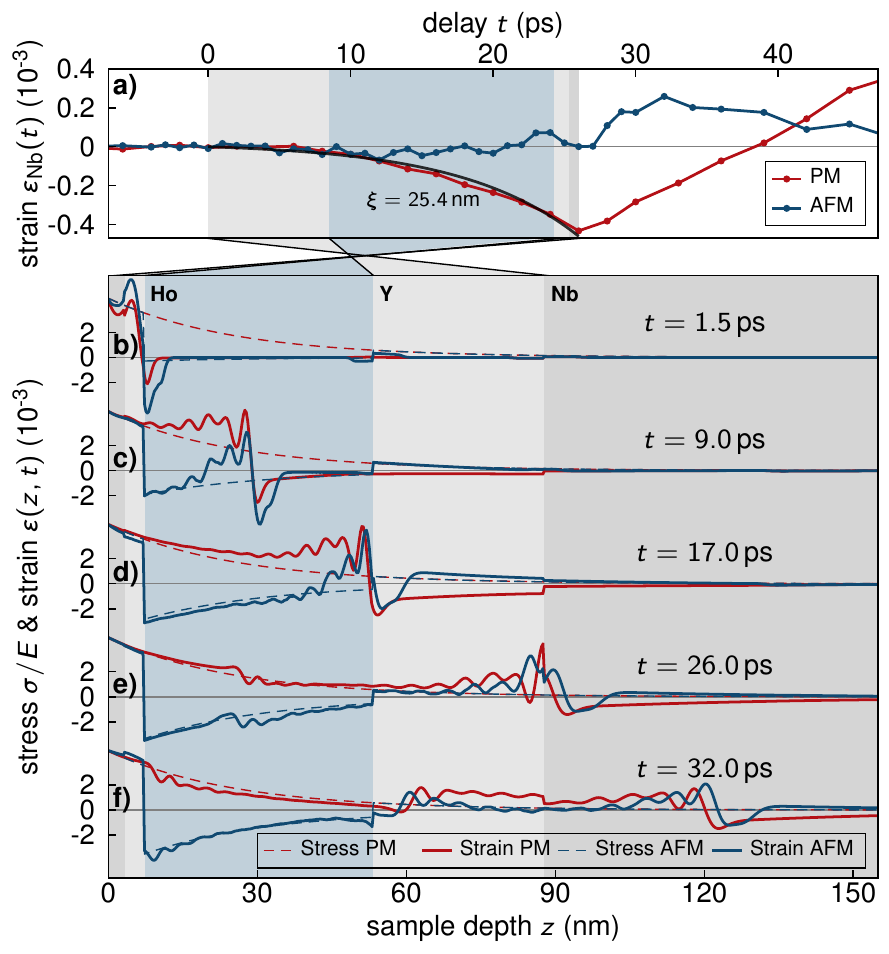}
  \caption{a) Zoom into the first 45\,ps of the Nb data from Fig.~\ref{fig:data}c). The black line shows an exponential fit to the strain in the PM phase, which is used to extract the driving stress via eq. 1. b-f) Stress (dashed) and strain profiles for selected delays in the heterostructure from a simulation. To avoid unimportant rapid oscillations in this graph, the temporal stress profile was smoothed by a 0.5 ps Gaussian function. The NTE stress in Ho rises with $\tau=12$\,ps.
  }
\label{fig:strain}
\end{figure}

In this paragraph we highlight the potential of UXRD for deriving the spatial form of the stress driving the observed strain.\cite{ruel2015,Pers2016} In Fig.~\ref{fig:strain}a) we analyze the UXRD data recorded for the Nb layer in the PM state of Ho, zooming into the pronounced compression signal of the Nb layer. The average strain $\varepsilon_\text{Nb}$ shows an increasing compression slowly starting at $t=0$ when the  bipolar strain wave starts entering the Nb layer. Neglecting sound velocity differences in the heterostructure, an exponential spatial stress profile generates a bipolar strain pulse with a compressive leading edge that has an exponential spatial dependence as well. \cite{schi2014c}
Thus, from the measured strain in a dedicated detection layer, we can show that the spatial stress profile in fact decays exponentially with the characteristic length scale $\xi=25$\,nm which matches the value calculated from the optical constants measured by ellipsometry. For direct comparison with the data we plotted in Fig.~\ref{fig:strain}a) the stress $\sigma_\text{PM}=\sigma_\text{0} \cdot e^{-z/\xi}$ onto the measured strain data using an appropriate scaling by Hooke's law. In the AFM phase (blue) the spin- and phonon stress contributions approximately cancel out  immediately after excitation.\\
\begin{figure}
   \centering
 \includegraphics[width = 8cm]{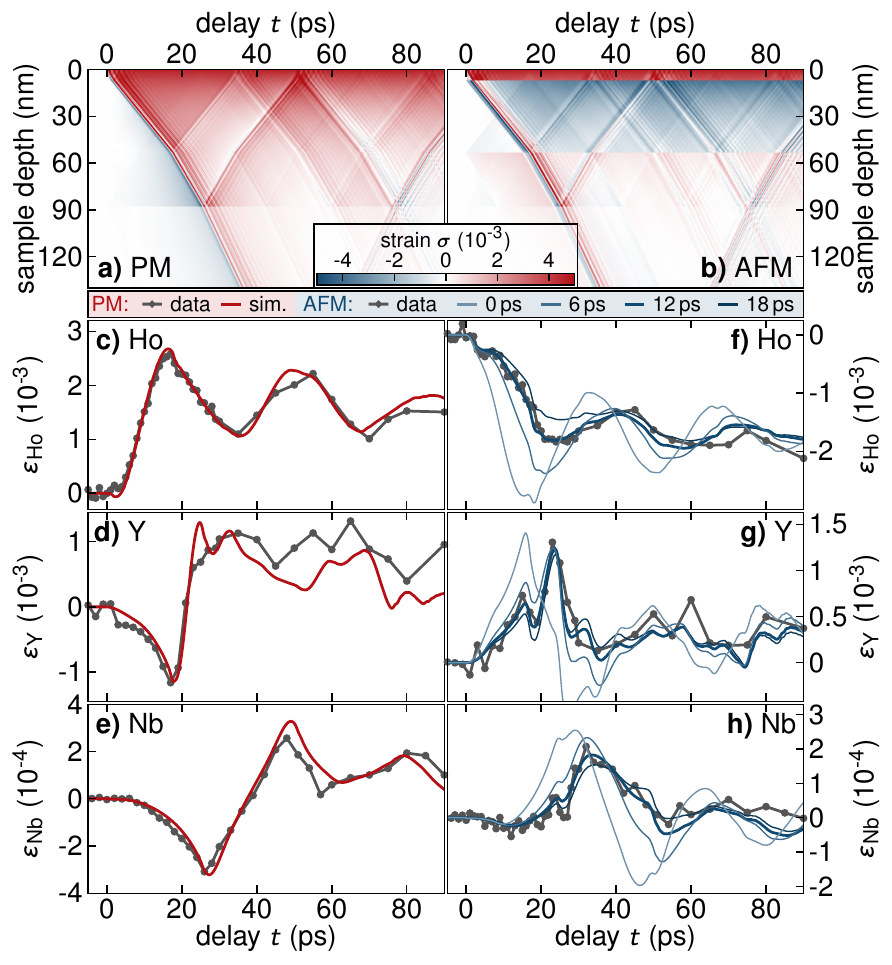}
  \caption{The simulated strain maps shows the travelling sound waves in the heterostructure in the a) PM and b) AFM phase. c-e) Simulated strain for the PM phase in the three layers (red) is compared to the data. f-h) Simulations for the AFM phase with varying time constant $\tau=0$, 6, 12 and 18\,ps. The best match to the data is obtained for $\tau=12$\,ps.}
\label{fig:sim}
\end{figure}
Now we discuss the numerical modelling of the observed strain that is necessary to derive the transient stress changing faster than the characteristic thickness modulation time of the layers. In the PM phase of Ho we calculate the transient strain (Fig.~\ref{fig:sim}a) by integrating the equation of motion for a linear masses-and-springs-model using the udkm1Dsim toolbox.\cite{schi2014b} The experimentally derived exponential spatial form of the stress serves as an input. Figs.~\ref{fig:sim}c,d,e) show  the excellent agreement of the simulations with the measured data for all three layers in the PM phase. 
For simplicity, we assume an instantaneous rise of the combined e-ph stress. Separate electron and phonon Gcs do not improve the fit in  Fig.~\ref{fig:sim}c,d,e). 
The position of the simulated Bragg peaks are obtained by fitting a Gaussian function as done in the experimental data analysis.
For illustration, Figs.~\ref{fig:strain}b-f) depict the transient strains leading to the fits in Fig.~4. Each figure contains also the transient spatial stress profiles (dashed) used as input for the simulation. 
After about 30\,ps the simulated strain profile is very close to the simple stress-strain relation, with only slight deviation due to residual hypersound waves.
When the negative stress induced by spin-disorder adds to the positve e-ph stress, the dynamics become richer and challenge the modelling. We model the transient stress $\sigma=\Gamma_\text{e-ph}\rho^Q_\text{e-ph}+\Gamma_\text{sp}\rho^Q_\text{sp}$ (Fig.~3b-f) from the balance of thermal energy densities in the combined e-ph system
 $\rho^Q_\text{e-ph}$ and in the spin system $\rho^Q_\text{s}$:
\begin{align}
\rho^Q_\text{e-ph}(t) &= \rho^Q_{\text{e-ph},\infty} + \rho^Q_\text{dyn} \cdot e^{-t/\tau}\label{eq3}\\
\rho^Q_\text{sp}(t) &= \rho^Q_\text{sp,0} + \rho^Q_\text{dyn} \cdot \left( 1 - e^{-t/\tau} \right)\label{eq4}
\end{align}
In the first 70\,ps in Fig.~\ref{fig:sim}f,g,h), we may disregard heat transport and assume local conversion of energy $\rho^Q_\text{e-ph}(t)$ to $\rho^Q_\text{sp}(t)$ with a coupling time $\tau$. In this model $\rho^Q_\text{sp,0}$ is an energy that is transferred to the spin-system quasi-instantaneously, while $\rho^Q_\text{dyn}$ is transferred from the e-ph system to the spins on the timescale $\tau$.
$\rho^Q_{\text{e-ph},\infty}$ is the heat remaining in the e-ph system until heat transport starts to become relevant. 
The partitioning into only two simulated heat reservoirs is chosen because $\rho^Q_\text{e-ph}$ and $\rho^Q_\text{sp}$ trigger competing expansive and contractive stresses, respectively. The essential fitting parameter is $\tau$, which we initially assumed to coincide with the slow timescale for demagnetization of 9\,ps observed by resonant x-ray scattering. \cite{rett2016} The fast time-scale is beyond the time-resolution we can extract from the lattice dynamics of layers with a thickness on the order of 50\,nm. We assume an instantaneous coupling of heat energy into the spin system inherent in eq.~\eqref{eq4}. Figs.~\ref{fig:sim}f-h) show the excellent agreement achieved simultaneously in the signals from all three layers for $\tau =12$\,ps. Especially the initial contraction of Ho and the delayed expansion of Nb are very sensitive to variations of $\tau$.

We would like to point out that the intense bipolar strain wave launched by the expansion of the capping layer at low temperatures evident from Fig.~4b) enhances our confidence in the model, since it is directly observed in the signal: The short dips in the Y strain (Fig.~4g) at 18\,ps and in the Nb strain (Fig.~4h) at 26\,ps indicate the arrival of the narrow compressive part of the bipolar strain pulse, independent of $\tau$. For $\tau=0$\,ps , the maximum in the Nb strain coincides with the minimum observed in the PM phase of Ho. 
The increasing delay of the maximum in Nb for increasing $\tau$ can be assigned to the delayed NTE stress induced by spin disordering. The blue line in Fig.~\ref{fig:strain}f) shows once more that the total strain essentially follows the NTE stress profile in Ho  (dashed). 
Variation of the spatial stress profile of the spin excitation, e.g. to a homogeneous demagnetization profile throughout the layer, has negligible effect on the Ho signal and does not improve the agreement with the Y and Nb signals.

Finally, we would like to discuss the observation of ultrafast negative stress on the sub-picosecond timescale in the context of NTE in general.
NTE requires a specific interaction, where the energy decreases with changes in the volume. \cite{barr1980,whit1962} For Ho the dominant interaction is the exchange interaction $J_{\text{exc}}$ and it has been shown earlier that the expansion coefficient $\alpha \sim  \partial J_{\text{exc}}/\partial c$ scales with the strain induced change of the exchange interaction $J_{\text{exc}}$.\cite{darn1963,Pytt1965} From a statistical physics perspective, the spin-entropy $S=k_\text{B} \ln(2J+1)$ must be dominated by the large angular momentum $J$ of the localized $4f$ moments. Hence, also the heat energy density $\rho^Q_{\text{sp}}$ associated with the spin disordering and the concomitant stress $\sigma(t) = \Gamma_{\text{sp}} \rho^Q_{\text{sp}}(t)$ is mainly connected to the localized spins. On the other hand, the RKKY interaction requires that the itinerant electrons mediate the coupling. It is not self-evident if a disordering of the optically excited itinerant electrons alone can explain the magnitude of the negative stress. In the system Ho, the recent work by Rettig \cite{rett2016} confirmed that both types of electrons disorder on the same timescale, however, it will be interesting to test the situation in other systems, such as Gd, where disparate timescales have been observed.\cite{frie2015}

In conclusion, we have reported that ultrafast laser-induced disordering of the spin system of Ho proceeds on two timescales and triggers NTE via ultrafast stress mediated by the exchange interaction. On the sub-picosecond timescale already nearly half of the negative stress is present and it fully balances the expansive stress from electrons and phonons. According to the ratio of Gcs $\Gamma_\text{e-ph}/\Gamma_\text{sp}=-0.2$   this balance implies that  20$\%$ of the energy absorbed in the Ho layer have excited the spin system. Within $\tau = 12$\,ps, the fraction of energy in the spin system rises to 40$\%$. If we consider the fact that for the ferromagnetic rare earth Gd, different timescales for disordering the localized and itinerant orbitals have been observed, it is not clear, on which timescale the stress should occur.
We believe that this study may trigger similar investigations in other systems with NTE, since it is not obvious that the thermodynamic relation predicting stress $\sigma = \Gamma\cdot \rho^Q$ proportional to the energy density in a subsystem will hold in time-dependent nonequilibrium situations and for any origin of the negative entropy-volume relation required for NTE.

\begin{acknowledgments}
We acknowledge the BMBF for the financial support
via 05K16IPA and the DFG via BA 2281/8-1 and BA 2281/11-1. We would like to thank J\"urgen Podschwadek for the MBE
sample preparation
\end{acknowledgments}



\begin{thebibliography}{42}%
\makeatletter
\providecommand \@ifxundefined [1]{%
 \@ifx{#1\undefined}
}%
\providecommand \@ifnum [1]{%
 \ifnum #1\expandafter \@firstoftwo
 \else \expandafter \@secondoftwo
 \fi
}%
\providecommand \@ifx [1]{%
 \ifx #1\expandafter \@firstoftwo
 \else \expandafter \@secondoftwo
 \fi
}%
\providecommand \natexlab [1]{#1}%
\providecommand \enquote  [1]{``#1''}%
\providecommand \bibnamefont  [1]{#1}%
\providecommand \bibfnamefont [1]{#1}%
\providecommand \citenamefont [1]{#1}%
\providecommand \href@noop [0]{\@secondoftwo}%
\providecommand \href [0]{\begingroup \@sanitize@url \@href}%
\providecommand \@href[1]{\@@startlink{#1}\@@href}%
\providecommand \@@href[1]{\endgroup#1\@@endlink}%
\providecommand \@sanitize@url [0]{\catcode `\\12\catcode `\$12\catcode
  `\&12\catcode `\#12\catcode `\^12\catcode `\_12\catcode `\%12\relax}%
\providecommand \@@startlink[1]{}%
\providecommand \@@endlink[0]{}%
\providecommand \url  [0]{\begingroup\@sanitize@url \@url }%
\providecommand \@url [1]{\endgroup\@href {#1}{\urlprefix }}%
\providecommand \urlprefix  [0]{URL }%
\providecommand \Eprint [0]{\href }%
\providecommand \doibase [0]{http://dx.doi.org/}%
\providecommand \selectlanguage [0]{\@gobble}%
\providecommand \bibinfo  [0]{\@secondoftwo}%
\providecommand \bibfield  [0]{\@secondoftwo}%
\providecommand \translation [1]{[#1]}%
\providecommand \BibitemOpen [0]{}%
\providecommand \bibitemStop [0]{}%
\providecommand \bibitemNoStop [0]{.\EOS\space}%
\providecommand \EOS [0]{\spacefactor3000\relax}%
\providecommand \BibitemShut  [1]{\csname bibitem#1\endcsname}%
\let\auto@bib@innerbib\@empty
\bibitem [{\citenamefont {Beaurepaire}\ \emph {et~al.}(1996)\citenamefont
  {Beaurepaire}, \citenamefont {Merle}, \citenamefont {Daunois},\ and\
  \citenamefont {Bigot}}]{beau1996a}%
  \BibitemOpen
  \bibfield  {author} {\bibinfo {author} {\bibfnamefont {E.}~\bibnamefont
  {Beaurepaire}}, \bibinfo {author} {\bibfnamefont {J.-C.}\ \bibnamefont
  {Merle}}, \bibinfo {author} {\bibfnamefont {A.}~\bibnamefont {Daunois}}, \
  and\ \bibinfo {author} {\bibfnamefont {J.-Y.}\ \bibnamefont {Bigot}},\
  }\bibfield  {title} {\enquote {\bibinfo {title} {{Ultrafast Spin Dynamics in
  Ferromagnetic Nickel}},}\ }\href {\doibase 10.1103/PhysRevLett.76.4250}
  {\bibfield  {journal} {\bibinfo  {journal} {Physical Review Letters}\
  }\textbf {\bibinfo {volume} {76}},\ \bibinfo {pages} {4250--4253} (\bibinfo
  {year} {1996})}\BibitemShut {NoStop}%
\bibitem [{\citenamefont {Bigot}\ \emph {et~al.}(2009)\citenamefont {Bigot},
  \citenamefont {Vomir},\ and\ \citenamefont {Beaurepaire}}]{bigo2009}%
  \BibitemOpen
  \bibfield  {author} {\bibinfo {author} {\bibfnamefont {Jean-Yves}\
  \bibnamefont {Bigot}}, \bibinfo {author} {\bibfnamefont {Mircea}\
  \bibnamefont {Vomir}}, \ and\ \bibinfo {author} {\bibfnamefont {Eric}\
  \bibnamefont {Beaurepaire}},\ }\bibfield  {title} {\enquote {\bibinfo {title}
  {{Coherent ultrafast magnetism induced by femtosecond laser pulses}},}\
  }\href {\doibase 10.1038/nphys1285} {\bibfield  {journal} {\bibinfo
  {journal} {Nature Physics}\ }\textbf {\bibinfo {volume} {5}},\ \bibinfo
  {pages} {515--520} (\bibinfo {year} {2009})}\BibitemShut {NoStop}%
\bibitem [{\citenamefont {Stamm}\ \emph {et~al.}(2007)\citenamefont {Stamm},
  \citenamefont {Kachel}, \citenamefont {Pontius}, \citenamefont {Mitzner},
  \citenamefont {Quast}, \citenamefont {Holldack}, \citenamefont {Khan},
  \citenamefont {Lupulescu}, \citenamefont {Aziz}, \citenamefont {Wietstruk},
  \citenamefont {D{\"{u}}rr},\ and\ \citenamefont {Eberhardt}}]{stam2007}%
  \BibitemOpen
  \bibfield  {author} {\bibinfo {author} {\bibfnamefont {C.}~\bibnamefont
  {Stamm}}, \bibinfo {author} {\bibfnamefont {T.}~\bibnamefont {Kachel}},
  \bibinfo {author} {\bibfnamefont {N.}~\bibnamefont {Pontius}}, \bibinfo
  {author} {\bibfnamefont {R.}~\bibnamefont {Mitzner}}, \bibinfo {author}
  {\bibfnamefont {T.}~\bibnamefont {Quast}}, \bibinfo {author} {\bibfnamefont
  {K.}~\bibnamefont {Holldack}}, \bibinfo {author} {\bibfnamefont
  {S.}~\bibnamefont {Khan}}, \bibinfo {author} {\bibfnamefont {C.}~\bibnamefont
  {Lupulescu}}, \bibinfo {author} {\bibfnamefont {E.~F.}\ \bibnamefont {Aziz}},
  \bibinfo {author} {\bibfnamefont {M.}~\bibnamefont {Wietstruk}}, \bibinfo
  {author} {\bibfnamefont {H.~A.}\ \bibnamefont {D{\"{u}}rr}}, \ and\ \bibinfo
  {author} {\bibfnamefont {W.}~\bibnamefont {Eberhardt}},\ }\bibfield  {title}
  {\enquote {\bibinfo {title} {{Femtosecond modification of electron
  localization and transfer of angular momentum in nickel}},}\ }\href {\doibase
  10.1038/nmat1985} {\bibfield  {journal} {\bibinfo  {journal} {Nature
  Materials}\ }\textbf {\bibinfo {volume} {6}},\ \bibinfo {pages} {740--743}
  (\bibinfo {year} {2007})}\BibitemShut {NoStop}%
\bibitem [{\citenamefont {{Korff Schmising}}\ \emph {et~al.}(2008)\citenamefont
  {{Korff Schmising}}, \citenamefont {Harpoeth}, \citenamefont {Zhavoronkov},
  \citenamefont {Ansari}, \citenamefont {Aku-Leh}, \citenamefont {Woerner},
  \citenamefont {Elsaesser}, \citenamefont {Bargheer}, \citenamefont
  {Schmidbauer}, \citenamefont {Vrejoiu}, \citenamefont {Hesse},\ and\
  \citenamefont {Alexe}}]{korf2008}%
  \BibitemOpen
  \bibfield  {author} {\bibinfo {author} {\bibfnamefont {C.~v.}\ \bibnamefont
  {{Korff Schmising}}}, \bibinfo {author} {\bibfnamefont {A.}~\bibnamefont
  {Harpoeth}}, \bibinfo {author} {\bibfnamefont {N.}~\bibnamefont
  {Zhavoronkov}}, \bibinfo {author} {\bibfnamefont {Z.}~\bibnamefont {Ansari}},
  \bibinfo {author} {\bibfnamefont {C.}~\bibnamefont {Aku-Leh}}, \bibinfo
  {author} {\bibfnamefont {M.}~\bibnamefont {Woerner}}, \bibinfo {author}
  {\bibfnamefont {T.}~\bibnamefont {Elsaesser}}, \bibinfo {author}
  {\bibfnamefont {M.}~\bibnamefont {Bargheer}}, \bibinfo {author}
  {\bibfnamefont {M.}~\bibnamefont {Schmidbauer}}, \bibinfo {author}
  {\bibfnamefont {I.}~\bibnamefont {Vrejoiu}}, \bibinfo {author} {\bibfnamefont
  {D.}~\bibnamefont {Hesse}}, \ and\ \bibinfo {author} {\bibfnamefont
  {M.}~\bibnamefont {Alexe}},\ }\bibfield  {title} {\enquote {\bibinfo {title}
  {{Ultrafast magnetostriction and phonon-mediated stress in a photoexcited
  ferromagnet}},}\ }\href {\doibase 10.1103/PhysRevB.78.060404} {\bibfield
  {journal} {\bibinfo  {journal} {Physical Review B}\ }\textbf {\bibinfo
  {volume} {78}},\ \bibinfo {pages} {060404} (\bibinfo {year}
  {2008})}\BibitemShut {NoStop}%
\bibitem [{\citenamefont {Reid}\ \emph {et~al.}(2018)\citenamefont {Reid},
  \citenamefont {Shen}, \citenamefont {Maldonado}, \citenamefont {Chase},
  \citenamefont {Jal}, \citenamefont {Granitzka}, \citenamefont {Carva},
  \citenamefont {Li}, \citenamefont {Li}, \citenamefont {Wu}, \citenamefont
  {Vecchione}, \citenamefont {Liu}, \citenamefont {Chen}, \citenamefont
  {Higley}, \citenamefont {Hartmann}, \citenamefont {Coffee}, \citenamefont
  {Wu}, \citenamefont {Dakovski}, \citenamefont {Schlotter}, \citenamefont
  {Ohldag}, \citenamefont {Takahashi}, \citenamefont {Mehta}, \citenamefont
  {Hellwig}, \citenamefont {Fry}, \citenamefont {Zhu}, \citenamefont {Cao},
  \citenamefont {Fullerton}, \citenamefont {St{\"{o}}hr}, \citenamefont
  {Oppeneer}, \citenamefont {Wang},\ and\ \citenamefont
  {D{\"{u}}rr}}]{reid2018}%
  \BibitemOpen
  \bibfield  {author} {\bibinfo {author} {\bibfnamefont {A.~H.}\ \bibnamefont
  {Reid}}, \bibinfo {author} {\bibfnamefont {X.}~\bibnamefont {Shen}}, \bibinfo
  {author} {\bibfnamefont {P.}~\bibnamefont {Maldonado}}, \bibinfo {author}
  {\bibfnamefont {T.}~\bibnamefont {Chase}}, \bibinfo {author} {\bibfnamefont
  {E.}~\bibnamefont {Jal}}, \bibinfo {author} {\bibfnamefont {P.~W.}\
  \bibnamefont {Granitzka}}, \bibinfo {author} {\bibfnamefont {K.}~\bibnamefont
  {Carva}}, \bibinfo {author} {\bibfnamefont {R.~K.}\ \bibnamefont {Li}},
  \bibinfo {author} {\bibfnamefont {J.}~\bibnamefont {Li}}, \bibinfo {author}
  {\bibfnamefont {L.}~\bibnamefont {Wu}}, \bibinfo {author} {\bibfnamefont
  {T.}~\bibnamefont {Vecchione}}, \bibinfo {author} {\bibfnamefont
  {T.}~\bibnamefont {Liu}}, \bibinfo {author} {\bibfnamefont {Z.}~\bibnamefont
  {Chen}}, \bibinfo {author} {\bibfnamefont {D.~J.}\ \bibnamefont {Higley}},
  \bibinfo {author} {\bibfnamefont {N.}~\bibnamefont {Hartmann}}, \bibinfo
  {author} {\bibfnamefont {R.}~\bibnamefont {Coffee}}, \bibinfo {author}
  {\bibfnamefont {J.}~\bibnamefont {Wu}}, \bibinfo {author} {\bibfnamefont
  {G.~L.}\ \bibnamefont {Dakovski}}, \bibinfo {author} {\bibfnamefont {W.~F.}\
  \bibnamefont {Schlotter}}, \bibinfo {author} {\bibfnamefont {H.}~\bibnamefont
  {Ohldag}}, \bibinfo {author} {\bibfnamefont {Y.~K.}\ \bibnamefont
  {Takahashi}}, \bibinfo {author} {\bibfnamefont {V.}~\bibnamefont {Mehta}},
  \bibinfo {author} {\bibfnamefont {O.}~\bibnamefont {Hellwig}}, \bibinfo
  {author} {\bibfnamefont {A.}~\bibnamefont {Fry}}, \bibinfo {author}
  {\bibfnamefont {Y.}~\bibnamefont {Zhu}}, \bibinfo {author} {\bibfnamefont
  {J.}~\bibnamefont {Cao}}, \bibinfo {author} {\bibfnamefont {E.~E.}\
  \bibnamefont {Fullerton}}, \bibinfo {author} {\bibfnamefont {J.}~\bibnamefont
  {St{\"{o}}hr}}, \bibinfo {author} {\bibfnamefont {P.~M.}\ \bibnamefont
  {Oppeneer}}, \bibinfo {author} {\bibfnamefont {X.~J.}\ \bibnamefont {Wang}},
  \ and\ \bibinfo {author} {\bibfnamefont {H.~A.}\ \bibnamefont {D{\"{u}}rr}},\
  }\bibfield  {title} {\enquote {\bibinfo {title} {{Beyond a phenomenological
  description of magnetostriction}},}\ }\href {\doibase
  10.1038/s41467-017-02730-7} {\bibfield  {journal} {\bibinfo  {journal}
  {Nature Communications}\ }\textbf {\bibinfo {volume} {9}},\ \bibinfo {pages}
  {388} (\bibinfo {year} {2018})}\BibitemShut {NoStop}%
\bibitem [{\citenamefont {Malinowski}\ \emph {et~al.}(2008)\citenamefont
  {Malinowski}, \citenamefont {{Dalla Longa}}, \citenamefont {Rietjens},
  \citenamefont {Paluskar}, \citenamefont {Huijink}, \citenamefont {Swagten},\
  and\ \citenamefont {Koopmans}}]{mali2008}%
  \BibitemOpen
  \bibfield  {author} {\bibinfo {author} {\bibfnamefont {G.}~\bibnamefont
  {Malinowski}}, \bibinfo {author} {\bibfnamefont {F.}~\bibnamefont {{Dalla
  Longa}}}, \bibinfo {author} {\bibfnamefont {J.~H.~H.}\ \bibnamefont
  {Rietjens}}, \bibinfo {author} {\bibfnamefont {P.~V.}\ \bibnamefont
  {Paluskar}}, \bibinfo {author} {\bibfnamefont {R.}~\bibnamefont {Huijink}},
  \bibinfo {author} {\bibfnamefont {H.~J.~M.}\ \bibnamefont {Swagten}}, \ and\
  \bibinfo {author} {\bibfnamefont {B.}~\bibnamefont {Koopmans}},\ }\bibfield
  {title} {\enquote {\bibinfo {title} {{Control of speed and efficiency of
  ultrafast demagnetization by direct transfer of spin angular momentum}},}\
  }\href {\doibase 10.1038/nphys1092} {\bibfield  {journal} {\bibinfo
  {journal} {Nature Physics}\ }\textbf {\bibinfo {volume} {4}},\ \bibinfo
  {pages} {855--858} (\bibinfo {year} {2008})}\BibitemShut {NoStop}%
\bibitem [{\citenamefont {Rudolf}\ \emph {et~al.}(2012)\citenamefont {Rudolf},
  \citenamefont {La-O-Vorakiat}, \citenamefont {Battiato}, \citenamefont
  {Adam}, \citenamefont {Shaw}, \citenamefont {Turgut}, \citenamefont
  {Maldonado}, \citenamefont {Mathias}, \citenamefont {Grychtol}, \citenamefont
  {Nembach}, \citenamefont {Silva}, \citenamefont {Aeschlimann}, \citenamefont
  {Kapteyn}, \citenamefont {Murnane}, \citenamefont {Schneider},\ and\
  \citenamefont {Oppeneer}}]{rudo2012}%
  \BibitemOpen
  \bibfield  {author} {\bibinfo {author} {\bibfnamefont {Dennis}\ \bibnamefont
  {Rudolf}}, \bibinfo {author} {\bibfnamefont {Chan}\ \bibnamefont
  {La-O-Vorakiat}}, \bibinfo {author} {\bibfnamefont {Marco}\ \bibnamefont
  {Battiato}}, \bibinfo {author} {\bibfnamefont {Roman}\ \bibnamefont {Adam}},
  \bibinfo {author} {\bibfnamefont {Justin~M.}\ \bibnamefont {Shaw}}, \bibinfo
  {author} {\bibfnamefont {Emrah}\ \bibnamefont {Turgut}}, \bibinfo {author}
  {\bibfnamefont {Pablo}\ \bibnamefont {Maldonado}}, \bibinfo {author}
  {\bibfnamefont {Stefan}\ \bibnamefont {Mathias}}, \bibinfo {author}
  {\bibfnamefont {Patrik}\ \bibnamefont {Grychtol}}, \bibinfo {author}
  {\bibfnamefont {Hans~T.}\ \bibnamefont {Nembach}}, \bibinfo {author}
  {\bibfnamefont {Thomas~J.}\ \bibnamefont {Silva}}, \bibinfo {author}
  {\bibfnamefont {Martin}\ \bibnamefont {Aeschlimann}}, \bibinfo {author}
  {\bibfnamefont {Henry~C.}\ \bibnamefont {Kapteyn}}, \bibinfo {author}
  {\bibfnamefont {Margaret~M.}\ \bibnamefont {Murnane}}, \bibinfo {author}
  {\bibfnamefont {Claus~M.}\ \bibnamefont {Schneider}}, \ and\ \bibinfo
  {author} {\bibfnamefont {Peter~M.}\ \bibnamefont {Oppeneer}},\ }\bibfield
  {title} {\enquote {\bibinfo {title} {{Ultrafast magnetization enhancement in
  metallic multilayers driven by superdiffusive spin current}},}\ }\href
  {\doibase 10.1038/ncomms2029} {\bibfield  {journal} {\bibinfo  {journal}
  {Nature Communications}\ }\textbf {\bibinfo {volume} {3}},\ \bibinfo {pages}
  {1037} (\bibinfo {year} {2012})}\BibitemShut {NoStop}%
\bibitem [{\citenamefont {Thomsen}\ \emph {et~al.}(1986)\citenamefont
  {Thomsen}, \citenamefont {Grahn}, \citenamefont {Maris},\ and\ \citenamefont
  {Tauc}}]{thom1986a}%
  \BibitemOpen
  \bibfield  {author} {\bibinfo {author} {\bibfnamefont {C.}~\bibnamefont
  {Thomsen}}, \bibinfo {author} {\bibfnamefont {H.~T.}\ \bibnamefont {Grahn}},
  \bibinfo {author} {\bibfnamefont {H.~J.}\ \bibnamefont {Maris}}, \ and\
  \bibinfo {author} {\bibfnamefont {J.}~\bibnamefont {Tauc}},\ }\bibfield
  {title} {\enquote {\bibinfo {title} {Surface generation and detection of
  phonons by picosecond light pulses},}\ }\href {\doibase
  10.1103/physrevb.34.4129} {\bibfield  {journal} {\bibinfo  {journal} {Phys.
  Rev. B}\ }\textbf {\bibinfo {volume} {34}},\ \bibinfo {pages} {4129--4138}
  (\bibinfo {year} {1986})}\BibitemShut {NoStop}%
\bibitem [{\citenamefont {Koopmans}\ \emph {et~al.}(2010)\citenamefont
  {Koopmans}, \citenamefont {Malinowski}, \citenamefont {{Dalla Longa}},
  \citenamefont {Steiauf}, \citenamefont {F{\"{a}}hnle}, \citenamefont {Roth},
  \citenamefont {Cinchetti},\ and\ \citenamefont {Aeschlimann}}]{koop2010a}%
  \BibitemOpen
  \bibfield  {author} {\bibinfo {author} {\bibfnamefont {B.}~\bibnamefont
  {Koopmans}}, \bibinfo {author} {\bibfnamefont {G.}~\bibnamefont
  {Malinowski}}, \bibinfo {author} {\bibfnamefont {F.}~\bibnamefont {{Dalla
  Longa}}}, \bibinfo {author} {\bibfnamefont {D.}~\bibnamefont {Steiauf}},
  \bibinfo {author} {\bibfnamefont {M.}~\bibnamefont {F{\"{a}}hnle}}, \bibinfo
  {author} {\bibfnamefont {T.}~\bibnamefont {Roth}}, \bibinfo {author}
  {\bibfnamefont {M.}~\bibnamefont {Cinchetti}}, \ and\ \bibinfo {author}
  {\bibfnamefont {M.}~\bibnamefont {Aeschlimann}},\ }\bibfield  {title}
  {\enquote {\bibinfo {title} {{Explaining the paradoxical diversity of
  ultrafast laser-induced demagnetization}},}\ }\href {\doibase
  10.1038/nmat2593} {\bibfield  {journal} {\bibinfo  {journal} {Nature
  Materials}\ }\textbf {\bibinfo {volume} {9}},\ \bibinfo {pages} {259--265}
  (\bibinfo {year} {2010})}\BibitemShut {NoStop}%
\bibitem [{\citenamefont {Hofherr}\ \emph {et~al.}(2017)\citenamefont
  {Hofherr}, \citenamefont {Maldonado}, \citenamefont {Schmitt}, \citenamefont
  {Berritta}, \citenamefont {Bierbrauer}, \citenamefont {Sadashivaiah},
  \citenamefont {Schellekens}, \citenamefont {Koopmans}, \citenamefont {Steil},
  \citenamefont {Cinchetti}, \citenamefont {Stadtm{\"{u}}ller}, \citenamefont
  {Oppeneer}, \citenamefont {Mathias},\ and\ \citenamefont
  {Aeschlimann}}]{hofh2017}%
  \BibitemOpen
  \bibfield  {author} {\bibinfo {author} {\bibfnamefont {M.}~\bibnamefont
  {Hofherr}}, \bibinfo {author} {\bibfnamefont {P.}~\bibnamefont {Maldonado}},
  \bibinfo {author} {\bibfnamefont {O.}~\bibnamefont {Schmitt}}, \bibinfo
  {author} {\bibfnamefont {M.}~\bibnamefont {Berritta}}, \bibinfo {author}
  {\bibfnamefont {U.}~\bibnamefont {Bierbrauer}}, \bibinfo {author}
  {\bibfnamefont {S.}~\bibnamefont {Sadashivaiah}}, \bibinfo {author}
  {\bibfnamefont {A.~J.}\ \bibnamefont {Schellekens}}, \bibinfo {author}
  {\bibfnamefont {B.}~\bibnamefont {Koopmans}}, \bibinfo {author}
  {\bibfnamefont {D.}~\bibnamefont {Steil}}, \bibinfo {author} {\bibfnamefont
  {M.}~\bibnamefont {Cinchetti}}, \bibinfo {author} {\bibfnamefont
  {B.}~\bibnamefont {Stadtm{\"{u}}ller}}, \bibinfo {author} {\bibfnamefont
  {P.~M.}\ \bibnamefont {Oppeneer}}, \bibinfo {author} {\bibfnamefont
  {S.}~\bibnamefont {Mathias}}, \ and\ \bibinfo {author} {\bibfnamefont
  {M.}~\bibnamefont {Aeschlimann}},\ }\bibfield  {title} {\enquote {\bibinfo
  {title} {{Speed and efficiency of femtosecond spin current injection into a
  nonmagnetic material}},}\ }\href {\doibase 10.1103/PhysRevB.96.100403}
  {\bibfield  {journal} {\bibinfo  {journal} {Physical Review B}\ }\textbf
  {\bibinfo {volume} {96}},\ \bibinfo {pages} {100403} (\bibinfo {year}
  {2017})}\BibitemShut {NoStop}%
\bibitem [{\citenamefont {Kimel}\ \emph {et~al.}(2005)\citenamefont {Kimel},
  \citenamefont {Kirilyuk}, \citenamefont {Usachev}, \citenamefont {Pisarev},
  \citenamefont {Balbashov},\ and\ \citenamefont {Rasing}}]{kime2005}%
  \BibitemOpen
  \bibfield  {author} {\bibinfo {author} {\bibfnamefont {A.~V.}\ \bibnamefont
  {Kimel}}, \bibinfo {author} {\bibfnamefont {A.}~\bibnamefont {Kirilyuk}},
  \bibinfo {author} {\bibfnamefont {P.~A.}\ \bibnamefont {Usachev}}, \bibinfo
  {author} {\bibfnamefont {R.~V.}\ \bibnamefont {Pisarev}}, \bibinfo {author}
  {\bibfnamefont {A.~M.}\ \bibnamefont {Balbashov}}, \ and\ \bibinfo {author}
  {\bibfnamefont {Th.}\ \bibnamefont {Rasing}},\ }\bibfield  {title} {\enquote
  {\bibinfo {title} {{Ultrafast non-thermal control of magnetization by
  instantaneous photomagnetic pulses}},}\ }\href {\doibase 10.1038/nature03564}
  {\bibfield  {journal} {\bibinfo  {journal} {Nature}\ }\textbf {\bibinfo
  {volume} {435}},\ \bibinfo {pages} {655--657} (\bibinfo {year}
  {2005})}\BibitemShut {NoStop}%
\bibitem [{\citenamefont {Kim}\ \emph {et~al.}(2012)\citenamefont {Kim},
  \citenamefont {Vomir},\ and\ \citenamefont {Bigot}}]{kim2012}%
  \BibitemOpen
  \bibfield  {author} {\bibinfo {author} {\bibfnamefont {Ji-Wan}\ \bibnamefont
  {Kim}}, \bibinfo {author} {\bibfnamefont {Mircea}\ \bibnamefont {Vomir}}, \
  and\ \bibinfo {author} {\bibfnamefont {Jean-Yves}\ \bibnamefont {Bigot}},\
  }\bibfield  {title} {\enquote {\bibinfo {title} {Ultrafast magnetoacoustics
  in nickel films},}\ }\href {\doibase 10.1103/PhysRevLett.109.166601}
  {\bibfield  {journal} {\bibinfo  {journal} {Phys. Rev. Lett.}\ }\textbf
  {\bibinfo {volume} {109}},\ \bibinfo {pages} {166601} (\bibinfo {year}
  {2012})}\BibitemShut {NoStop}%
\bibitem [{\citenamefont {Rettig}\ \emph {et~al.}(2016)\citenamefont {Rettig},
  \citenamefont {Dornes}, \citenamefont {Thielemann-K{\"{u}}hn}, \citenamefont
  {Pontius}, \citenamefont {Zabel}, \citenamefont {Schlagel}, \citenamefont
  {Lograsso}, \citenamefont {Chollet}, \citenamefont {Robert}, \citenamefont
  {Sikorski}, \citenamefont {Song}, \citenamefont {Glownia}, \citenamefont
  {Sch{\"{u}}{\ss}ler-Langeheine}, \citenamefont {Johnson},\ and\ \citenamefont
  {Staub}}]{rett2016}%
  \BibitemOpen
  \bibfield  {author} {\bibinfo {author} {\bibfnamefont {L.}~\bibnamefont
  {Rettig}}, \bibinfo {author} {\bibfnamefont {C.}~\bibnamefont {Dornes}},
  \bibinfo {author} {\bibfnamefont {N.}~\bibnamefont {Thielemann-K{\"{u}}hn}},
  \bibinfo {author} {\bibfnamefont {N.}~\bibnamefont {Pontius}}, \bibinfo
  {author} {\bibfnamefont {H.}~\bibnamefont {Zabel}}, \bibinfo {author}
  {\bibfnamefont {D.~L.}\ \bibnamefont {Schlagel}}, \bibinfo {author}
  {\bibfnamefont {T.~A.}\ \bibnamefont {Lograsso}}, \bibinfo {author}
  {\bibfnamefont {M.}~\bibnamefont {Chollet}}, \bibinfo {author} {\bibfnamefont
  {A.}~\bibnamefont {Robert}}, \bibinfo {author} {\bibfnamefont
  {M.}~\bibnamefont {Sikorski}}, \bibinfo {author} {\bibfnamefont
  {S.}~\bibnamefont {Song}}, \bibinfo {author} {\bibfnamefont {J.~M.}\
  \bibnamefont {Glownia}}, \bibinfo {author} {\bibfnamefont {C.}~\bibnamefont
  {Sch{\"{u}}{\ss}ler-Langeheine}}, \bibinfo {author} {\bibfnamefont {S.~L.}\
  \bibnamefont {Johnson}}, \ and\ \bibinfo {author} {\bibfnamefont
  {U.}~\bibnamefont {Staub}},\ }\bibfield  {title} {\enquote {\bibinfo {title}
  {{Itinerant and Localized Magnetization Dynamics in Antiferromagnetic Ho}},}\
  }\href {\doibase 10.1103/PhysRevLett.116.257202} {\bibfield  {journal}
  {\bibinfo  {journal} {Physical Review Letters}\ }\textbf {\bibinfo {volume}
  {116}},\ \bibinfo {pages} {257202} (\bibinfo {year} {2016})}\BibitemShut
  {NoStop}%
\bibitem [{\citenamefont {Thielemann-K{\"{u}}hn}\ \emph
  {et~al.}(2017)\citenamefont {Thielemann-K{\"{u}}hn}, \citenamefont {Schick},
  \citenamefont {Pontius}, \citenamefont {Trabant}, \citenamefont {Mitzner},
  \citenamefont {Holldack}, \citenamefont {Zabel}, \citenamefont
  {F{\"{o}}hlisch},\ and\ \citenamefont
  {Sch{\"{u}}{\ss}ler-Langeheine}}]{thie2017}%
  \BibitemOpen
  \bibfield  {author} {\bibinfo {author} {\bibfnamefont {Nele}\ \bibnamefont
  {Thielemann-K{\"{u}}hn}}, \bibinfo {author} {\bibfnamefont {Daniel}\
  \bibnamefont {Schick}}, \bibinfo {author} {\bibfnamefont {Niko}\ \bibnamefont
  {Pontius}}, \bibinfo {author} {\bibfnamefont {Christoph}\ \bibnamefont
  {Trabant}}, \bibinfo {author} {\bibfnamefont {Rolf}\ \bibnamefont {Mitzner}},
  \bibinfo {author} {\bibfnamefont {Karsten}\ \bibnamefont {Holldack}},
  \bibinfo {author} {\bibfnamefont {Hartmut}\ \bibnamefont {Zabel}}, \bibinfo
  {author} {\bibfnamefont {Alexander}\ \bibnamefont {F{\"{o}}hlisch}}, \ and\
  \bibinfo {author} {\bibfnamefont {Christian}\ \bibnamefont
  {Sch{\"{u}}{\ss}ler-Langeheine}},\ }\bibfield  {title} {\enquote {\bibinfo
  {title} {{Ultrafast and Energy-Efficient Quenching of Spin Order:
  Antiferromagnetism Beats Ferromagnetism}},}\ }\href {\doibase
  10.1103/PhysRevLett.119.197202} {\bibfield  {journal} {\bibinfo  {journal}
  {Physical Review Letters}\ }\textbf {\bibinfo {volume} {119}},\ \bibinfo
  {pages} {197202} (\bibinfo {year} {2017})}\BibitemShut {NoStop}%
\bibitem [{\citenamefont {Pfau}\ \emph {et~al.}(2012)\citenamefont {Pfau},
  \citenamefont {Schaffert}, \citenamefont {M{\"{u}}ller}, \citenamefont
  {Gutt}, \citenamefont {Al-Shemmary}, \citenamefont {B{\"{u}}ttner},
  \citenamefont {Delaunay}, \citenamefont {D{\"{u}}sterer}, \citenamefont
  {Flewett}, \citenamefont {Fr{\"{o}}mter}, \citenamefont {Geilhufe},
  \citenamefont {Guehrs}, \citenamefont {G{\"{u}}nther}, \citenamefont
  {Hawaldar}, \citenamefont {Hille}, \citenamefont {Jaouen}, \citenamefont
  {Kobs}, \citenamefont {Li}, \citenamefont {Mohanty}, \citenamefont {Redlin},
  \citenamefont {Schlotter}, \citenamefont {Stickler}, \citenamefont {Treusch},
  \citenamefont {Vodungbo}, \citenamefont {Kl{\"{a}}ui}, \citenamefont {Oepen},
  \citenamefont {L{\"{u}}ning}, \citenamefont {Gr{\"{u}}bel},\ and\
  \citenamefont {Eisebitt}}]{pfau2012}%
  \BibitemOpen
  \bibfield  {author} {\bibinfo {author} {\bibfnamefont {B.}~\bibnamefont
  {Pfau}}, \bibinfo {author} {\bibfnamefont {S.}~\bibnamefont {Schaffert}},
  \bibinfo {author} {\bibfnamefont {L.}~\bibnamefont {M{\"{u}}ller}}, \bibinfo
  {author} {\bibfnamefont {C.}~\bibnamefont {Gutt}}, \bibinfo {author}
  {\bibfnamefont {A.}~\bibnamefont {Al-Shemmary}}, \bibinfo {author}
  {\bibfnamefont {F.}~\bibnamefont {B{\"{u}}ttner}}, \bibinfo {author}
  {\bibfnamefont {R.}~\bibnamefont {Delaunay}}, \bibinfo {author}
  {\bibfnamefont {S.}~\bibnamefont {D{\"{u}}sterer}}, \bibinfo {author}
  {\bibfnamefont {S.}~\bibnamefont {Flewett}}, \bibinfo {author} {\bibfnamefont
  {R.}~\bibnamefont {Fr{\"{o}}mter}}, \bibinfo {author} {\bibfnamefont
  {J.}~\bibnamefont {Geilhufe}}, \bibinfo {author} {\bibfnamefont
  {E.}~\bibnamefont {Guehrs}}, \bibinfo {author} {\bibfnamefont {C.M.}\
  \bibnamefont {G{\"{u}}nther}}, \bibinfo {author} {\bibfnamefont
  {R.}~\bibnamefont {Hawaldar}}, \bibinfo {author} {\bibfnamefont
  {M.}~\bibnamefont {Hille}}, \bibinfo {author} {\bibfnamefont
  {N.}~\bibnamefont {Jaouen}}, \bibinfo {author} {\bibfnamefont
  {A.}~\bibnamefont {Kobs}}, \bibinfo {author} {\bibfnamefont {K.}~\bibnamefont
  {Li}}, \bibinfo {author} {\bibfnamefont {J.}~\bibnamefont {Mohanty}},
  \bibinfo {author} {\bibfnamefont {H.}~\bibnamefont {Redlin}}, \bibinfo
  {author} {\bibfnamefont {W.F.}\ \bibnamefont {Schlotter}}, \bibinfo {author}
  {\bibfnamefont {D.}~\bibnamefont {Stickler}}, \bibinfo {author}
  {\bibfnamefont {R.}~\bibnamefont {Treusch}}, \bibinfo {author} {\bibfnamefont
  {B.}~\bibnamefont {Vodungbo}}, \bibinfo {author} {\bibfnamefont
  {M.}~\bibnamefont {Kl{\"{a}}ui}}, \bibinfo {author} {\bibfnamefont {H.P.}\
  \bibnamefont {Oepen}}, \bibinfo {author} {\bibfnamefont {J.}~\bibnamefont
  {L{\"{u}}ning}}, \bibinfo {author} {\bibfnamefont {G.}~\bibnamefont
  {Gr{\"{u}}bel}}, \ and\ \bibinfo {author} {\bibfnamefont {S.}~\bibnamefont
  {Eisebitt}},\ }\bibfield  {title} {\enquote {\bibinfo {title} {{Ultrafast
  optical demagnetization manipulates nanoscale spin structure in domain
  walls}},}\ }\href {\doibase 10.1038/ncomms2108} {\bibfield  {journal}
  {\bibinfo  {journal} {Nature Communications}\ }\textbf {\bibinfo {volume}
  {3}},\ \bibinfo {pages} {1100} (\bibinfo {year} {2012})}\BibitemShut
  {NoStop}%
\bibitem [{\citenamefont {{Doerr *}}\ \emph {et~al.}(2005)\citenamefont {{Doerr
  *}}, \citenamefont {Rotter},\ and\ \citenamefont {Lindbaum}}]{doer2005}%
  \BibitemOpen
  \bibfield  {author} {\bibinfo {author} {\bibfnamefont {M.}~\bibnamefont
  {{Doerr *}}}, \bibinfo {author} {\bibfnamefont {M.}~\bibnamefont {Rotter}}, \
  and\ \bibinfo {author} {\bibfnamefont {A.}~\bibnamefont {Lindbaum}},\
  }\bibfield  {title} {\enquote {\bibinfo {title} {{Magnetostriction in
  rare-earth based antiferromagnets}},}\ }\href {\doibase
  10.1080/00018730500037264} {\bibfield  {journal} {\bibinfo  {journal}
  {Advances in Physics}\ }\textbf {\bibinfo {volume} {54}},\ \bibinfo {pages}
  {1--66} (\bibinfo {year} {2005})}\BibitemShut {NoStop}%
\bibitem [{\citenamefont {Ruderman}\ and\ \citenamefont
  {Kittel}(1954)}]{rude1954}%
  \BibitemOpen
  \bibfield  {author} {\bibinfo {author} {\bibfnamefont {M.~A.}\ \bibnamefont
  {Ruderman}}\ and\ \bibinfo {author} {\bibfnamefont {C.}~\bibnamefont
  {Kittel}},\ }\bibfield  {title} {\enquote {\bibinfo {title} {{Indirect
  Exchange Coupling of Nuclear Magnetic Moments by Conduction Electrons}},}\
  }\href {\doibase 10.1103/PhysRev.96.99} {\bibfield  {journal} {\bibinfo
  {journal} {Physical Review}\ }\textbf {\bibinfo {volume} {96}},\ \bibinfo
  {pages} {99--102} (\bibinfo {year} {1954})}\BibitemShut {NoStop}%
\bibitem [{\citenamefont {Darnell}(1963)}]{darn1963}%
  \BibitemOpen
  \bibfield  {author} {\bibinfo {author} {\bibfnamefont {F.~J.}\ \bibnamefont
  {Darnell}},\ }\bibfield  {title} {\enquote {\bibinfo {title}
  {{Magnetostriction in Dysprosium and Terbium}},}\ }\href {\doibase
  10.1103/PhysRev.132.128} {\bibfield  {journal} {\bibinfo  {journal} {Physical
  Review}\ }\textbf {\bibinfo {volume} {132}},\ \bibinfo {pages} {128--133}
  (\bibinfo {year} {1963})}\BibitemShut {NoStop}%
\bibitem [{\citenamefont {Koc}\ \emph {et~al.}(2017)\citenamefont {Koc},
  \citenamefont {Reinhardt}, \citenamefont {von Reppert}, \citenamefont
  {R{\"{o}}ssle}, \citenamefont {Leitenberger}, \citenamefont {Gleich},
  \citenamefont {Weinelt}, \citenamefont {Zamponi},\ and\ \citenamefont
  {Bargheer}}]{koc2017}%
  \BibitemOpen
  \bibfield  {author} {\bibinfo {author} {\bibfnamefont {A}~\bibnamefont
  {Koc}}, \bibinfo {author} {\bibfnamefont {M}~\bibnamefont {Reinhardt}},
  \bibinfo {author} {\bibfnamefont {A}~\bibnamefont {von Reppert}}, \bibinfo
  {author} {\bibfnamefont {M}~\bibnamefont {R{\"{o}}ssle}}, \bibinfo {author}
  {\bibfnamefont {W}~\bibnamefont {Leitenberger}}, \bibinfo {author}
  {\bibfnamefont {M}~\bibnamefont {Gleich}}, \bibinfo {author} {\bibfnamefont
  {M}~\bibnamefont {Weinelt}}, \bibinfo {author} {\bibfnamefont
  {F}~\bibnamefont {Zamponi}}, \ and\ \bibinfo {author} {\bibfnamefont
  {M}~\bibnamefont {Bargheer}},\ }\bibfield  {title} {\enquote {\bibinfo
  {title} {{Grueneisen-approach for the experimental determination of transient
  spin and phonon energies from ultrafast x-ray diffraction data:
  gadolinium}},}\ }\href {\doibase 10.1088/1361-648X/aa7187} {\bibfield
  {journal} {\bibinfo  {journal} {Journal of Physics: Condensed Matter}\
  }\textbf {\bibinfo {volume} {29}},\ \bibinfo {pages} {264001} (\bibinfo
  {year} {2017})}\BibitemShut {NoStop}%
\bibitem [{\citenamefont {von Reppert}\ \emph {et~al.}(2016)\citenamefont {von
  Reppert}, \citenamefont {Pudell}, \citenamefont {Koc}, \citenamefont
  {Reinhardt}, \citenamefont {Leitenberger}, \citenamefont {Dumesnil},
  \citenamefont {Zamponi},\ and\ \citenamefont {Bargheer}}]{repp2016}%
  \BibitemOpen
  \bibfield  {author} {\bibinfo {author} {\bibfnamefont {A.}~\bibnamefont {von
  Reppert}}, \bibinfo {author} {\bibfnamefont {J.}~\bibnamefont {Pudell}},
  \bibinfo {author} {\bibfnamefont {A.}~\bibnamefont {Koc}}, \bibinfo {author}
  {\bibfnamefont {M.}~\bibnamefont {Reinhardt}}, \bibinfo {author}
  {\bibfnamefont {W.}~\bibnamefont {Leitenberger}}, \bibinfo {author}
  {\bibfnamefont {K.}~\bibnamefont {Dumesnil}}, \bibinfo {author}
  {\bibfnamefont {F.}~\bibnamefont {Zamponi}}, \ and\ \bibinfo {author}
  {\bibfnamefont {M.}~\bibnamefont {Bargheer}},\ }\bibfield  {title} {\enquote
  {\bibinfo {title} {{Persistent nonequilibrium dynamics of the thermal
  energies in the spin and phonon systems of an antiferromagnet}},}\ }\href
  {\doibase 10.1063/1.4961253} {\bibfield  {journal} {\bibinfo  {journal}
  {Structural Dynamics}\ }\textbf {\bibinfo {volume} {3}},\ \bibinfo {pages}
  {054302} (\bibinfo {year} {2016})}\BibitemShut {NoStop}%
\bibitem [{\citenamefont {Quirin}\ \emph {et~al.}(2012)\citenamefont {Quirin},
  \citenamefont {Vattilana}, \citenamefont {Shymanovich}, \citenamefont
  {El-Kamhawy}, \citenamefont {Tarasevitch}, \citenamefont {Hohlfeld},
  \citenamefont {von~der Linde},\ and\ \citenamefont
  {Sokolowski-Tinten}}]{quir2012}%
  \BibitemOpen
  \bibfield  {author} {\bibinfo {author} {\bibfnamefont {Florian}\ \bibnamefont
  {Quirin}}, \bibinfo {author} {\bibfnamefont {Michael}\ \bibnamefont
  {Vattilana}}, \bibinfo {author} {\bibfnamefont {Uladzimir}\ \bibnamefont
  {Shymanovich}}, \bibinfo {author} {\bibfnamefont {Abd-Elmoniem}\ \bibnamefont
  {El-Kamhawy}}, \bibinfo {author} {\bibfnamefont {Alexander}\ \bibnamefont
  {Tarasevitch}}, \bibinfo {author} {\bibfnamefont {Julius}\ \bibnamefont
  {Hohlfeld}}, \bibinfo {author} {\bibfnamefont {Dietrich}\ \bibnamefont
  {von~der Linde}}, \ and\ \bibinfo {author} {\bibfnamefont {Klaus}\
  \bibnamefont {Sokolowski-Tinten}},\ }\bibfield  {title} {\enquote {\bibinfo
  {title} {{Structural dynamics in FeRh during a laser-induced metamagnetic
  phase transition}},}\ }\href {\doibase 10.1103/PhysRevB.85.020103} {\bibfield
   {journal} {\bibinfo  {journal} {Physical Review B}\ }\textbf {\bibinfo
  {volume} {85}},\ \bibinfo {pages} {020103} (\bibinfo {year}
  {2012})}\BibitemShut {NoStop}%
\bibitem [{\citenamefont {Wang}\ \emph {et~al.}(2008)\citenamefont {Wang},
  \citenamefont {Nie}, \citenamefont {Li}, \citenamefont {Clinite},
  \citenamefont {Wartenbe}, \citenamefont {Martin}, \citenamefont {Liang},\
  and\ \citenamefont {Cao}}]{wang2008}%
  \BibitemOpen
  \bibfield  {author} {\bibinfo {author} {\bibfnamefont {Xuan}\ \bibnamefont
  {Wang}}, \bibinfo {author} {\bibfnamefont {Shouhua}\ \bibnamefont {Nie}},
  \bibinfo {author} {\bibfnamefont {Junjie}\ \bibnamefont {Li}}, \bibinfo
  {author} {\bibfnamefont {Richard}\ \bibnamefont {Clinite}}, \bibinfo {author}
  {\bibfnamefont {Mark}\ \bibnamefont {Wartenbe}}, \bibinfo {author}
  {\bibfnamefont {Marcia}\ \bibnamefont {Martin}}, \bibinfo {author}
  {\bibfnamefont {Wenxi}\ \bibnamefont {Liang}}, \ and\ \bibinfo {author}
  {\bibfnamefont {Jianming}\ \bibnamefont {Cao}},\ }\bibfield  {title}
  {\enquote {\bibinfo {title} {{Electronic Gr{\"{u}}neisen parameter and
  thermal expansion in ferromagnetic transition metal}},}\ }\href {\doibase
  10.1063/1.2902170} {\bibfield  {journal} {\bibinfo  {journal} {Applied
  Physics Letters}\ }\textbf {\bibinfo {volume} {92}},\ \bibinfo {pages}
  {121918} (\bibinfo {year} {2008})}\BibitemShut {NoStop}%
\bibitem [{\citenamefont {Wang}\ \emph {et~al.}(2010)\citenamefont {Wang},
  \citenamefont {Nie}, \citenamefont {Li}, \citenamefont {Clinite},
  \citenamefont {Clark},\ and\ \citenamefont {Cao}}]{wang2010}%
  \BibitemOpen
  \bibfield  {author} {\bibinfo {author} {\bibfnamefont {Xuan}\ \bibnamefont
  {Wang}}, \bibinfo {author} {\bibfnamefont {Shouhua}\ \bibnamefont {Nie}},
  \bibinfo {author} {\bibfnamefont {Junjie}\ \bibnamefont {Li}}, \bibinfo
  {author} {\bibfnamefont {Richard}\ \bibnamefont {Clinite}}, \bibinfo {author}
  {\bibfnamefont {John~Edward}\ \bibnamefont {Clark}}, \ and\ \bibinfo {author}
  {\bibfnamefont {Jianming}\ \bibnamefont {Cao}},\ }\bibfield  {title}
  {\enquote {\bibinfo {title} {{Temperature dependence of electron-phonon
  thermalization and its correlation to ultrafast magnetism}},}\ }\href
  {\doibase 10.1103/PhysRevB.81.220301} {\bibfield  {journal} {\bibinfo
  {journal} {Physical Review B}\ }\textbf {\bibinfo {volume} {81}},\ \bibinfo
  {pages} {220301} (\bibinfo {year} {2010})}\BibitemShut {NoStop}%
\bibitem [{\citenamefont {Henighan}\ \emph {et~al.}(2016)\citenamefont
  {Henighan}, \citenamefont {Trigo}, \citenamefont {Bonetti}, \citenamefont
  {Granitzka}, \citenamefont {Higley}, \citenamefont {Chen}, \citenamefont
  {Jiang}, \citenamefont {Kukreja}, \citenamefont {Gray}, \citenamefont {Reid},
  \citenamefont {Jal}, \citenamefont {Hoffmann}, \citenamefont {Kozina},
  \citenamefont {Song}, \citenamefont {Chollet}, \citenamefont {Zhu},
  \citenamefont {Xu}, \citenamefont {Jeong}, \citenamefont {Carva},
  \citenamefont {Maldonado}, \citenamefont {Oppeneer}, \citenamefont {Samant},
  \citenamefont {Parkin}, \citenamefont {Reis},\ and\ \citenamefont
  {D{\"{u}}rr}}]{heni2016}%
  \BibitemOpen
  \bibfield  {author} {\bibinfo {author} {\bibfnamefont {T.}~\bibnamefont
  {Henighan}}, \bibinfo {author} {\bibfnamefont {M.}~\bibnamefont {Trigo}},
  \bibinfo {author} {\bibfnamefont {S.}~\bibnamefont {Bonetti}}, \bibinfo
  {author} {\bibfnamefont {P.}~\bibnamefont {Granitzka}}, \bibinfo {author}
  {\bibfnamefont {D.}~\bibnamefont {Higley}}, \bibinfo {author} {\bibfnamefont
  {Z.}~\bibnamefont {Chen}}, \bibinfo {author} {\bibfnamefont {M.~P.}\
  \bibnamefont {Jiang}}, \bibinfo {author} {\bibfnamefont {R.}~\bibnamefont
  {Kukreja}}, \bibinfo {author} {\bibfnamefont {A.}~\bibnamefont {Gray}},
  \bibinfo {author} {\bibfnamefont {A.~H.}\ \bibnamefont {Reid}}, \bibinfo
  {author} {\bibfnamefont {E.}~\bibnamefont {Jal}}, \bibinfo {author}
  {\bibfnamefont {M.~C.}\ \bibnamefont {Hoffmann}}, \bibinfo {author}
  {\bibfnamefont {M.}~\bibnamefont {Kozina}}, \bibinfo {author} {\bibfnamefont
  {S.}~\bibnamefont {Song}}, \bibinfo {author} {\bibfnamefont {M.}~\bibnamefont
  {Chollet}}, \bibinfo {author} {\bibfnamefont {D.}~\bibnamefont {Zhu}},
  \bibinfo {author} {\bibfnamefont {P.~F.}\ \bibnamefont {Xu}}, \bibinfo
  {author} {\bibfnamefont {J.}~\bibnamefont {Jeong}}, \bibinfo {author}
  {\bibfnamefont {K.}~\bibnamefont {Carva}}, \bibinfo {author} {\bibfnamefont
  {P.}~\bibnamefont {Maldonado}}, \bibinfo {author} {\bibfnamefont {P.~M.}\
  \bibnamefont {Oppeneer}}, \bibinfo {author} {\bibfnamefont {M.~G.}\
  \bibnamefont {Samant}}, \bibinfo {author} {\bibfnamefont {S.~S.~P.}\
  \bibnamefont {Parkin}}, \bibinfo {author} {\bibfnamefont {D.~A.}\
  \bibnamefont {Reis}}, \ and\ \bibinfo {author} {\bibfnamefont {H.~A.}\
  \bibnamefont {D{\"{u}}rr}},\ }\bibfield  {title} {\enquote {\bibinfo {title}
  {{Generation mechanism of terahertz coherent acoustic phonons in Fe}},}\
  }\href {\doibase 10.1103/PhysRevB.93.220301} {\bibfield  {journal} {\bibinfo
  {journal} {Physical Review B}\ }\textbf {\bibinfo {volume} {93}},\ \bibinfo
  {pages} {220301} (\bibinfo {year} {2016})}\BibitemShut {NoStop}%
\bibitem [{\citenamefont {Lin}\ \emph {et~al.}(2008)\citenamefont {Lin},
  \citenamefont {Zhigilei},\ and\ \citenamefont {Celli}}]{lin2008a}%
  \BibitemOpen
  \bibfield  {author} {\bibinfo {author} {\bibfnamefont {Zhibin}\ \bibnamefont
  {Lin}}, \bibinfo {author} {\bibfnamefont {Leonid~V.}\ \bibnamefont
  {Zhigilei}}, \ and\ \bibinfo {author} {\bibfnamefont {Vittorio}\ \bibnamefont
  {Celli}},\ }\bibfield  {title} {\enquote {\bibinfo {title} {{Electron-phonon
  coupling and electron heat capacity of metals under conditions of strong
  electron-phonon nonequilibrium}},}\ }\href {\doibase
  10.1103/PhysRevB.77.075133} {\bibfield  {journal} {\bibinfo  {journal}
  {Physical Review B}\ }\textbf {\bibinfo {volume} {77}},\ \bibinfo {pages}
  {075133} (\bibinfo {year} {2008})}\BibitemShut {NoStop}%
\bibitem [{\citenamefont {Maldonado}\ \emph {et~al.}(2017)\citenamefont
  {Maldonado}, \citenamefont {Carva}, \citenamefont {Flammer},\ and\
  \citenamefont {Oppeneer}}]{mald2017}%
  \BibitemOpen
  \bibfield  {author} {\bibinfo {author} {\bibfnamefont {Pablo}\ \bibnamefont
  {Maldonado}}, \bibinfo {author} {\bibfnamefont {Karel}\ \bibnamefont
  {Carva}}, \bibinfo {author} {\bibfnamefont {Martina}\ \bibnamefont
  {Flammer}}, \ and\ \bibinfo {author} {\bibfnamefont {Peter~M.}\ \bibnamefont
  {Oppeneer}},\ }\bibfield  {title} {\enquote {\bibinfo {title} {{Theory of
  out-of-equilibrium ultrafast relaxation dynamics in metals}},}\ }\href
  {\doibase 10.1103/PhysRevB.96.174439} {\bibfield  {journal} {\bibinfo
  {journal} {Physical Review B}\ }\textbf {\bibinfo {volume} {96}},\ \bibinfo
  {pages} {174439} (\bibinfo {year} {2017})}\BibitemShut {NoStop}%
\bibitem [{\citenamefont {Barrera}\ \emph {et~al.}(2005)\citenamefont
  {Barrera}, \citenamefont {Bruno}, \citenamefont {Barron},\ and\ \citenamefont
  {Allan}}]{barr2005}%
  \BibitemOpen
  \bibfield  {author} {\bibinfo {author} {\bibfnamefont {G~D}\ \bibnamefont
  {Barrera}}, \bibinfo {author} {\bibfnamefont {J~A~O}\ \bibnamefont {Bruno}},
  \bibinfo {author} {\bibfnamefont {T~H~K}\ \bibnamefont {Barron}}, \ and\
  \bibinfo {author} {\bibfnamefont {N~L}\ \bibnamefont {Allan}},\ }\bibfield
  {title} {\enquote {\bibinfo {title} {{Negative thermal expansion}},}\ }\href
  {\doibase 10.1088/0953-8984/17/4/R03} {\bibfield  {journal} {\bibinfo
  {journal} {Journal of Physics: Condensed Matter}\ }\textbf {\bibinfo {volume}
  {17}},\ \bibinfo {pages} {R217--R252} (\bibinfo {year} {2005})}\BibitemShut
  {NoStop}%
\bibitem [{\citenamefont {Nicoul}\ \emph {et~al.}(2011)\citenamefont {Nicoul},
  \citenamefont {Shymanovich}, \citenamefont {Tarasevitch}, \citenamefont
  {von~der Linde},\ and\ \citenamefont {Sokolowski-Tinten}}]{nico2011a}%
  \BibitemOpen
  \bibfield  {author} {\bibinfo {author} {\bibfnamefont {Matthieu}\
  \bibnamefont {Nicoul}}, \bibinfo {author} {\bibfnamefont {Uladzimir}\
  \bibnamefont {Shymanovich}}, \bibinfo {author} {\bibfnamefont {Alexander}\
  \bibnamefont {Tarasevitch}}, \bibinfo {author} {\bibfnamefont {Dietrich}\
  \bibnamefont {von~der Linde}}, \ and\ \bibinfo {author} {\bibfnamefont
  {Klaus}\ \bibnamefont {Sokolowski-Tinten}},\ }\bibfield  {title} {\enquote
  {\bibinfo {title} {{Picosecond acoustic response of a laser-heated gold-film
  studied with time-resolved x-ray diffraction}},}\ }\href {\doibase
  10.1063/1.3584864} {\bibfield  {journal} {\bibinfo  {journal} {Applied
  Physics Letters}\ }\textbf {\bibinfo {volume} {98}},\ \bibinfo {pages}
  {191902} (\bibinfo {year} {2011})}\BibitemShut {NoStop}%
\bibitem [{\citenamefont {White}(1962)}]{whit1962}%
  \BibitemOpen
  \bibfield  {author} {\bibinfo {author} {\bibfnamefont {G.K.}\ \bibnamefont
  {White}},\ }\bibfield  {title} {\enquote {\bibinfo {title} {{Thermal
  expansion at low temperatures— V. Dilute alloys of manganese in copper}},}\
  }\href {\doibase 10.1016/0022-3697(62)90077-X} {\bibfield  {journal}
  {\bibinfo  {journal} {Journal of Physics and Chemistry of Solids}\ }\textbf
  {\bibinfo {volume} {23}},\ \bibinfo {pages} {169--171} (\bibinfo {year}
  {1962})}\BibitemShut {NoStop}%
\bibitem [{\citenamefont {Barron}\ \emph {et~al.}(1980)\citenamefont {Barron},
  \citenamefont {Collins},\ and\ \citenamefont {White}}]{barr1980}%
  \BibitemOpen
  \bibfield  {author} {\bibinfo {author} {\bibfnamefont {T.H.K.}\ \bibnamefont
  {Barron}}, \bibinfo {author} {\bibfnamefont {J.G.}\ \bibnamefont {Collins}},
  \ and\ \bibinfo {author} {\bibfnamefont {G.K.}\ \bibnamefont {White}},\
  }\bibfield  {title} {\enquote {\bibinfo {title} {{Thermal expansion of solids
  at low temperatures}},}\ }\href {\doibase 10.1080/00018738000101426}
  {\bibfield  {journal} {\bibinfo  {journal} {Advances in Physics}\ }\textbf
  {\bibinfo {volume} {29}},\ \bibinfo {pages} {609--730} (\bibinfo {year}
  {1980})}\BibitemShut {NoStop}%
\bibitem [{\citenamefont {White}(1989)}]{whit1989}%
  \BibitemOpen
  \bibfield  {author} {\bibinfo {author} {\bibfnamefont {G~K}\ \bibnamefont
  {White}},\ }\bibfield  {title} {\enquote {\bibinfo {title} {{Phase
  transitions and the thermal expansion of holmium}},}\ }\href {\doibase
  10.1088/0953-8984/1/39/009} {\bibfield  {journal} {\bibinfo  {journal}
  {Journal of Physics: Condensed Matter}\ }\textbf {\bibinfo {volume} {1}},\
  \bibinfo {pages} {6987--6992} (\bibinfo {year} {1989})}\BibitemShut {NoStop}%
\bibitem [{\citenamefont {Schick}\ \emph
  {et~al.}(2014{\natexlab{a}})\citenamefont {Schick}, \citenamefont {Herzog},
  \citenamefont {Bojahr}, \citenamefont {Leitenberger}, \citenamefont
  {Hertwig}, \citenamefont {Shayduk},\ and\ \citenamefont
  {Bargheer}}]{schi2014c}%
  \BibitemOpen
  \bibfield  {author} {\bibinfo {author} {\bibfnamefont {D.}~\bibnamefont
  {Schick}}, \bibinfo {author} {\bibfnamefont {M.}~\bibnamefont {Herzog}},
  \bibinfo {author} {\bibfnamefont {A.}~\bibnamefont {Bojahr}}, \bibinfo
  {author} {\bibfnamefont {W.}~\bibnamefont {Leitenberger}}, \bibinfo {author}
  {\bibfnamefont {A.}~\bibnamefont {Hertwig}}, \bibinfo {author} {\bibfnamefont
  {R.}~\bibnamefont {Shayduk}}, \ and\ \bibinfo {author} {\bibfnamefont
  {M.}~\bibnamefont {Bargheer}},\ }\bibfield  {title} {\enquote {\bibinfo
  {title} {Ultrafast lattice response of photoexcited thin films studied by
  x-ray diffraction},}\ }\href {\doibase 10.1063/1.4901228} {\bibfield
  {journal} {\bibinfo  {journal} {Structural Dynamics}\ }\textbf {\bibinfo
  {volume} {1}} (\bibinfo {year} {2014}{\natexlab{a}}),\
  10.1063/1.4901228}\BibitemShut {NoStop}%
\bibitem [{\citenamefont {van~der Veen}\ \emph {et~al.}(2013)\citenamefont
  {van~der Veen}, \citenamefont {Kwon}, \citenamefont {Tissot}, \citenamefont
  {Hauser},\ and\ \citenamefont {Zewail}}]{van2013}%
  \BibitemOpen
  \bibfield  {author} {\bibinfo {author} {\bibfnamefont {Renske~M.}\
  \bibnamefont {van~der Veen}}, \bibinfo {author} {\bibfnamefont {Oh-Hoon}\
  \bibnamefont {Kwon}}, \bibinfo {author} {\bibfnamefont {Antoine}\
  \bibnamefont {Tissot}}, \bibinfo {author} {\bibfnamefont {Andreas}\
  \bibnamefont {Hauser}}, \ and\ \bibinfo {author} {\bibfnamefont {Ahmed~H.}\
  \bibnamefont {Zewail}},\ }\bibfield  {title} {\enquote {\bibinfo {title}
  {{Single-nanoparticle phase transitions visualized by four-dimensional
  electron microscopy}},}\ }\href {\doibase 10.1038/nchem.1622} {\bibfield
  {journal} {\bibinfo  {journal} {Nature Chemistry}\ }\textbf {\bibinfo
  {volume} {5}},\ \bibinfo {pages} {395--402} (\bibinfo {year}
  {2013})}\BibitemShut {NoStop}%
\bibitem [{\citenamefont {Ernst}\ \emph {et~al.}(1998)\citenamefont {Ernst},
  \citenamefont {Broholm}, \citenamefont {Kowach},\ and\ \citenamefont
  {Ramirez}}]{erns1998}%
  \BibitemOpen
  \bibfield  {author} {\bibinfo {author} {\bibfnamefont {G}~\bibnamefont
  {Ernst}}, \bibinfo {author} {\bibfnamefont {C}~\bibnamefont {Broholm}},
  \bibinfo {author} {\bibfnamefont {GR}~\bibnamefont {Kowach}}, \ and\ \bibinfo
  {author} {\bibfnamefont {AP}~\bibnamefont {Ramirez}},\ }\bibfield  {title}
  {\enquote {\bibinfo {title} {Phonon density of states and negative thermal
  expansion in zrw 2 o 8},}\ }\href@noop {} {\bibfield  {journal} {\bibinfo
  {journal} {Nature}\ }\textbf {\bibinfo {volume} {396}},\ \bibinfo {pages}
  {147} (\bibinfo {year} {1998})}\BibitemShut {NoStop}%
\bibitem [{\citenamefont {Chen}\ \emph {et~al.}(2013)\citenamefont {Chen},
  \citenamefont {Fan}, \citenamefont {Ren}, \citenamefont {Pan}, \citenamefont
  {Deng}, \citenamefont {Yu},\ and\ \citenamefont {Xing}}]{chen2013}%
  \BibitemOpen
  \bibfield  {author} {\bibinfo {author} {\bibfnamefont {Jun}\ \bibnamefont
  {Chen}}, \bibinfo {author} {\bibfnamefont {Longlong}\ \bibnamefont {Fan}},
  \bibinfo {author} {\bibfnamefont {Yang}\ \bibnamefont {Ren}}, \bibinfo
  {author} {\bibfnamefont {Zhao}\ \bibnamefont {Pan}}, \bibinfo {author}
  {\bibfnamefont {Jinxia}\ \bibnamefont {Deng}}, \bibinfo {author}
  {\bibfnamefont {Ranbo}\ \bibnamefont {Yu}}, \ and\ \bibinfo {author}
  {\bibfnamefont {Xianran}\ \bibnamefont {Xing}},\ }\bibfield  {title}
  {\enquote {\bibinfo {title} {Unusual transformation from strong negative to
  positive thermal expansion in
  ${\mathrm{pbtio}}_{3}\mathrm{\text{\ensuremath{-}}}{\mathrm{bifeo}}_{3}$
  perovskite},}\ }\href {\doibase 10.1103/PhysRevLett.110.115901} {\bibfield
  {journal} {\bibinfo  {journal} {Phys. Rev. Lett.}\ }\textbf {\bibinfo
  {volume} {110}},\ \bibinfo {pages} {115901} (\bibinfo {year}
  {2013})}\BibitemShut {NoStop}%
\bibitem [{\citenamefont {Khmelevskyi}\ \emph {et~al.}(2003)\citenamefont
  {Khmelevskyi}, \citenamefont {Turek},\ and\ \citenamefont {Mohn}}]{khme2003}%
  \BibitemOpen
  \bibfield  {author} {\bibinfo {author} {\bibfnamefont {S.}~\bibnamefont
  {Khmelevskyi}}, \bibinfo {author} {\bibfnamefont {I.}~\bibnamefont {Turek}},
  \ and\ \bibinfo {author} {\bibfnamefont {P.}~\bibnamefont {Mohn}},\
  }\bibfield  {title} {\enquote {\bibinfo {title} {{Large Negative Magnetic
  Contribution to the Thermal Expansion in Iron-Platinum Alloys: Quantitative
  Theory of the Invar Effect}},}\ }\href {\doibase
  10.1103/PhysRevLett.91.037201} {\bibfield  {journal} {\bibinfo  {journal}
  {Physical Review Letters}\ }\textbf {\bibinfo {volume} {91}},\ \bibinfo
  {pages} {037201} (\bibinfo {year} {2003})}\BibitemShut {NoStop}%
\bibitem [{\citenamefont {Schick}\ \emph {et~al.}(2012)\citenamefont {Schick},
  \citenamefont {Bojahr}, \citenamefont {Herzog}, \citenamefont {Schmising},
  \citenamefont {Shayduk}, \citenamefont {Leitenberger}, \citenamefont {Gaal},\
  and\ \citenamefont {Bargheer}}]{schi2012}%
  \BibitemOpen
  \bibfield  {author} {\bibinfo {author} {\bibfnamefont {D.}~\bibnamefont
  {Schick}}, \bibinfo {author} {\bibfnamefont {A.}~\bibnamefont {Bojahr}},
  \bibinfo {author} {\bibfnamefont {M.}~\bibnamefont {Herzog}}, \bibinfo
  {author} {\bibfnamefont {C.~von~Korff}\ \bibnamefont {Schmising}}, \bibinfo
  {author} {\bibfnamefont {R.}~\bibnamefont {Shayduk}}, \bibinfo {author}
  {\bibfnamefont {W.}~\bibnamefont {Leitenberger}}, \bibinfo {author}
  {\bibfnamefont {P.}~\bibnamefont {Gaal}}, \ and\ \bibinfo {author}
  {\bibfnamefont {M.}~\bibnamefont {Bargheer}},\ }\bibfield  {title} {\enquote
  {\bibinfo {title} {{Normalization schemes for ultrafast x-ray diffraction
  using a table-top laser-driven plasma source}},}\ }\href {\doibase
  10.1063/1.3681254} {\bibfield  {journal} {\bibinfo  {journal} {Review of
  Scientific Instruments}\ }\textbf {\bibinfo {volume} {83}},\ \bibinfo {pages}
  {025104} (\bibinfo {year} {2012})}\BibitemShut {NoStop}%
\bibitem [{\citenamefont {Ruello}\ and\ \citenamefont
  {Gusev}(2015)}]{ruel2015}%
  \BibitemOpen
  \bibfield  {author} {\bibinfo {author} {\bibfnamefont {Pascal}\ \bibnamefont
  {Ruello}}\ and\ \bibinfo {author} {\bibfnamefont {Vitalyi~E}\ \bibnamefont
  {Gusev}},\ }\bibfield  {title} {\enquote {\bibinfo {title} {{Physical
  mechanisms of coherent acoustic phonons generation by ultrafast laser
  action}},}\ }\href {\doibase 10.1016/j.ultras.2014.06.004} {\bibfield
  {journal} {\bibinfo  {journal} {Ultrasonics}\ }\textbf {\bibinfo {volume}
  {56}},\ \bibinfo {pages} {21--35} (\bibinfo {year} {2015})}\BibitemShut
  {NoStop}%
\bibitem [{\citenamefont {Persson}\ \emph {et~al.}(2016)\citenamefont
  {Persson}, \citenamefont {Jarnac}, \citenamefont {Wang}, \citenamefont
  {Enquist}, \citenamefont {Jurgilaitis},\ and\ \citenamefont
  {Larsson}}]{Pers2016}%
  \BibitemOpen
  \bibfield  {author} {\bibinfo {author} {\bibfnamefont {A.~I.~H.}\
  \bibnamefont {Persson}}, \bibinfo {author} {\bibfnamefont {A.}~\bibnamefont
  {Jarnac}}, \bibinfo {author} {\bibfnamefont {Xiaocui}\ \bibnamefont {Wang}},
  \bibinfo {author} {\bibfnamefont {H.}~\bibnamefont {Enquist}}, \bibinfo
  {author} {\bibfnamefont {A.}~\bibnamefont {Jurgilaitis}}, \ and\ \bibinfo
  {author} {\bibfnamefont {J.}~\bibnamefont {Larsson}},\ }\bibfield  {title}
  {\enquote {\bibinfo {title} {Studies of electron diffusion in photo-excited
  ni using time-resolved x-ray diffraction},}\ }\href {\doibase
  10.1063/1.4967470} {\bibfield  {journal} {\bibinfo  {journal} {Applied
  Physics Letters}\ }\textbf {\bibinfo {volume} {109}},\ \bibinfo {pages}
  {203115} (\bibinfo {year} {2016})},\ \Eprint
  {http://arxiv.org/abs/https://doi.org/10.1063/1.4967470}
  {https://doi.org/10.1063/1.4967470} \BibitemShut {NoStop}%
\bibitem [{\citenamefont {Schick}\ \emph
  {et~al.}(2014{\natexlab{b}})\citenamefont {Schick}, \citenamefont {Bojahr},
  \citenamefont {Herzog}, \citenamefont {Shayduk}, \citenamefont {{von Korff
  Schmising}},\ and\ \citenamefont {Bargheer}}]{schi2014b}%
  \BibitemOpen
  \bibfield  {author} {\bibinfo {author} {\bibfnamefont {D.}~\bibnamefont
  {Schick}}, \bibinfo {author} {\bibfnamefont {A.}~\bibnamefont {Bojahr}},
  \bibinfo {author} {\bibfnamefont {M.}~\bibnamefont {Herzog}}, \bibinfo
  {author} {\bibfnamefont {R.}~\bibnamefont {Shayduk}}, \bibinfo {author}
  {\bibfnamefont {C.}~\bibnamefont {{von Korff Schmising}}}, \ and\ \bibinfo
  {author} {\bibfnamefont {M.}~\bibnamefont {Bargheer}},\ }\bibfield  {title}
  {\enquote {\bibinfo {title} {{udkm1Dsim—A simulation toolkit for 1D
  ultrafast dynamics in condensed matter}},}\ }\href {\doibase
  10.1016/J.CPC.2013.10.009} {\bibfield  {journal} {\bibinfo  {journal}
  {Computer Physics Communications}\ }\textbf {\bibinfo {volume} {185}},\
  \bibinfo {pages} {651--660} (\bibinfo {year}
  {2014}{\natexlab{b}})}\BibitemShut {NoStop}%
\bibitem [{\citenamefont {Pytte}(1965)}]{Pytt1965}%
  \BibitemOpen
  \bibfield  {author} {\bibinfo {author} {\bibfnamefont {Erling}\ \bibnamefont
  {Pytte}},\ }\bibfield  {title} {\enquote {\bibinfo {title} {Spin-phonon
  interactions in a heisenberg ferromagnet},}\ }\href {\doibase
  http://dx.doi.org/10.1016/0003-4916(65)90139-9} {\bibfield  {journal}
  {\bibinfo  {journal} {Annals of Physics}\ }\textbf {\bibinfo {volume} {32}},\
  \bibinfo {pages} {377 -- 403} (\bibinfo {year} {1965})}\BibitemShut {NoStop}%
\bibitem [{\citenamefont {Frietsch}\ \emph {et~al.}(2015)\citenamefont
  {Frietsch}, \citenamefont {Bowlan}, \citenamefont {Carley}, \citenamefont
  {Teichmann}, \citenamefont {Wienholdt}, \citenamefont {Hinzke}, \citenamefont
  {Nowak}, \citenamefont {Carva}, \citenamefont {Oppeneer},\ and\ \citenamefont
  {Weinelt}}]{frie2015}%
  \BibitemOpen
  \bibfield  {author} {\bibinfo {author} {\bibfnamefont {B.}~\bibnamefont
  {Frietsch}}, \bibinfo {author} {\bibfnamefont {J.}~\bibnamefont {Bowlan}},
  \bibinfo {author} {\bibfnamefont {R.}~\bibnamefont {Carley}}, \bibinfo
  {author} {\bibfnamefont {M.}~\bibnamefont {Teichmann}}, \bibinfo {author}
  {\bibfnamefont {S.}~\bibnamefont {Wienholdt}}, \bibinfo {author}
  {\bibfnamefont {D.}~\bibnamefont {Hinzke}}, \bibinfo {author} {\bibfnamefont
  {U.}~\bibnamefont {Nowak}}, \bibinfo {author} {\bibfnamefont
  {K.}~\bibnamefont {Carva}}, \bibinfo {author} {\bibfnamefont {P.~M.}\
  \bibnamefont {Oppeneer}}, \ and\ \bibinfo {author} {\bibfnamefont
  {M.}~\bibnamefont {Weinelt}},\ }\bibfield  {title} {\enquote {\bibinfo
  {title} {{Disparate ultrafast dynamics of itinerant and localized magnetic
  moments in gadolinium metal}},}\ }\href {\doibase 10.1038/ncomms9262}
  {\bibfield  {journal} {\bibinfo  {journal} {Nature Communications}\ }\textbf
  {\bibinfo {volume} {6}},\ \bibinfo {pages} {8262} (\bibinfo {year}
  {2015})}\BibitemShut {NoStop}%
\end{thebibliography}
%

\end{document}